\documentclass[twocolumn,apj,numberedappendix,appendixfloats]{openjournal}

\usepackage{newtxtext,newtxmath,amsmath,amssymb}
\usepackage[T1]{fontenc}
\usepackage{graphicx}
\usepackage{enumitem}
\usepackage{xcolor}
\usepackage{hyperref}
\usepackage{gensymb}
\usepackage{CJK}
\usepackage{upgreek}
\usepackage{tabularx}
\usepackage{orcidlink}
\hypersetup{
    unicode, 
    colorlinks=true,
    linkcolor=linkcolor,
    citecolor=linkcolor,
    filecolor=linkcolor,
    urlcolor=linkcolor,
}
\definecolor{linkcolor}{rgb}{0.0,0.3,0.5}

\definecolor{M1}{HTML}{000000} 
\definecolor{M2}{HTML}{1F77B4} 
\definecolor{M3}{HTML}{FF7F0E} 
\definecolor{M4}{HTML}{9467BD} 
\definecolor{M5}{HTML}{D62728} 
\definecolor{M6}{HTML}{E377C2} 

\begin{document}
\title{Reading Between the Lines:\\How Much Information Remains After Fitting a Stellar Spectrum?}

\author{\vspace{-1.5cm}Sven Buder$^{1,*}$\orcidlink{0000-0002-4031-8553}}
\author{Tomasz Różánski$^{1}$\orcidlink{0000-0002-5819-3023}}
\affiliation{$^{1}$Research School of Astronomy and Astrophysics, Australian National University, Canberra, ACT 2611, Australia}

\thanks{E-mail: sven.buder@anu.edu.au}
\thanks{$^*$Australian Research Council DECRA Fellow}

\begin{abstract} 
What if we could model 100\,\% of a stellar spectrum? Modern surveys infer stellar properties by comparing observed spectra with synthetic models, yet even the best fits leave structured residuals. We analyse 7087 GALAH survey spectra and ask what information remains after subtracting each best-fitting model. Across 13927 wavelength pixels, an average of 4.5\,\% residuals exceed $3\sigma$, far above 0.3\,\% expected from Gaussian noise. We map the amplitude, significance, sign, and recurrence of these residuals and identify 380 significant regions. Comparison with stellar atlases, line lists, diffuse interstellar bands, and sky diagnostics links most regions to plausible causes: 62\,\% are candidate line-strength or line-profile mismatches, while others trace missing or unidentified lines, interstellar absorption, sky emission, or broader reduction structure. We then train data-driven models on the residual spectra themselves. Residual-only models recover $T_\mathrm{eff}$ to $80-150\,\mathrm{K}$ and $[\mathrm{Fe/H}]$ to $0.059-0.086\,\mathrm{dex}$, depending on sample quality, and retain weaker, partly indirect information about elemental abundances. Extinction-related coefficients recover the strongest known interstellar features, whereas chromospheric activity leaves plausible Balmer-line signatures but is not recovered robustly out of sample. Most strikingly, the residual flux in 6\,\% of wavelength pixels that were synthesised but excluded from the GALAH DR4 optimisation recovers stellar labels almost as well as the full residual spectrum and better than the remaining 94\,\% alone. Residuals are therefore not merely imperfections to discard: they map missing line data, incomplete physics, unfitted astrophysical components, and observational systematics. Reading between the lines turns one spectroscopist's trash into another one's treasure, and provides a practical route towards extracting more of the information already present in stellar spectra.
\end{abstract}


\maketitle

\section{Introduction}
\label{sec:introduction}

Stellar spectra are among the most information-rich observables in astrophysics. They encode atmospheric parameters, elemental abundances, line-formation physics, stellar activity, and non-stellar absorption. Large spectroscopic surveys use this information to reconstruct the chemical and dynamical evolution of the Milky Way \citep{FreemanBlandHawthorn2002, BlandHawthorn_Gerhard2016} and other galaxies \citep{VandeVen2025}. High-precision spectra are also central to exoplanet science, where planetary signals must be separated from stellar, telluric, and instrumental variability \citep[e.g.][]{Fischer2016}. Their scientific return therefore depends not only on the number of observed stars, but also on how completely the information in each spectrum is extracted.

This raises a deliberately provocative question: what if we could fit 100\% of a stellar spectrum? In practice, every stellar spectroscopic analysis leaves residuals between the observed spectrum and the best-fitting model. These residuals are often treated as a quality-control diagnostic or as a nuisance to be masked, but they may also encode astrophysical, observational, and modelling information that the adopted spectral model has not captured. This is not only a conceptual issue: in the Galactic Archaeology with HERMES (GALAH) survey, one of the large-scale stellar spectroscopic surveys of the Milky Way, machine-learning interpolation made it computationally feasible to increase the fraction of the spectrum used for stellar parameter and abundance fitting from typically $\sim20\%$ in Data Release 3 \citep[DR3,][]{Buder2021} to $87\%$ in Data Release 4 \citep[DR4,][]{Buder2025b}. However, even this modern synthesis could only use up to $94\%$ of the spectrum outside the typically neglected telluric O$_2$ Fraunhofer A-band region, because some observed spectral regions still could not be described reliably by the model \citep[][Sec.~4.3.2]{Buder2025b}. The central premise of this work is that the residual spectrum itself is a useful scientific object: it records where and how the model differs from the photons.

The modelling of stellar spectra has advanced substantially over the last decade \citep{Nissen2018,Jofre2019}. Modern survey pipelines increasingly infer many stellar labels simultaneously across broad wavelength ranges \citep[see Tab.~1 of][]{Ness2026b}, either through direct comparison with synthetic spectra or through flexible data-driven and emulator-based approaches \citep[e.g.][]{Ness2015,Casey2016,Ting2019}. Data-driven methods now support spectral generation, label inference, calibration, denoising, component separation, and residual analysis \citep[see][for a review]{Ness2026b}. Their learned spectral derivatives can recover known line sensitivities and reveal previously unrecognised abundance-sensitive features \citep{Buder2018}.

Despite this progress, no survey analysis reproduces every observed pixel at the level implied by its uncertainties. Residuals can reflect uncertain atomic or molecular data, missing blends or hyperfine structure \citep{Heiter2021}, continuum normalisation \citep{Jofre2017}, fixed abundance or isotopic assumptions \citep[e.g.][]{Coelho2007,Yong2004,Masseron2021}, incomplete stellar physics \citep{Amarsi2018,Masseron2021,Lind2024,Bergemann2026}, activity \citep{Lisogorskyi2019,Spina2020,Wilson2023b,Nordlander2024}, and interstellar, telluric, sky, or instrumental contamination \citep{Saydjari2023,Casey2026}. Unresolved multiplicity introduces a further contribution that we bracket out of this single-star audit \citep{ElBadry2018,Merle2024}. Such mismatches can nevertheless reveal usable information: C$_2$ and CN features initially considered inaccessible to GALAH \citep{DeSilva2015} were first identified in carbon-enhanced stars \citep{Cotar2019b} and subsequently enabled C and N measurements for a much broader sample \citep{Buder2025b}.

An important limitation is the completeness and accuracy of the underlying atomic and molecular data. Modern spectroscopic analyses rely on line lists that combine laboratory measurements, theoretical calculations, and astrophysical adjustments, but many transitions still lack experimentally validated wavelengths, oscillator strengths, isotopic shifts, hyperfine structure, or broadening parameters \citep{Heiter2015b}. As survey data quality improves, systematic residuals around known transitions, unidentified absorption features, and line-profile mismatches can therefore serve as observational diagnostics of where improved laboratory spectroscopy, molecular data, and atomic-structure calculations are most urgently needed.

Residual and component-separation methods have already extracted extinction, diffuse interstellar bands, sky emission, and instrumental structure from large stellar surveys, including LAMOST, GALAH, and APOGEE \citep{Ho2017,Buder2018,Vogrincic2023, McKinnon2024}. More general frameworks such as \textsc{MADGICS} explicitly separate stellar, dust, sky, instrumental, and residual components \citep{Saydjari2023}. These studies demonstrate that spectral mismatches can contain astrophysically useful information from several distinct sources. Here, we take a complementary approach: after subtracting the best-fitting physical synthesis, we retain the residuals in the stellar rest frame and ask what they reveal about fitted stellar labels, abundance patterns, extinction, activity, and model limitations. Our aim is to diagnose and guide improvements to the physical model, rather than to replace it.

GALAH is particularly well suited to such an audit because it provides observed spectra, best-fitting synthetic spectra, flux uncertainties, stellar parameters, elemental abundances, extinction estimates, quality information, and residual-based diagnostic products for a large, homogeneous optical data set \citep{DeSilva2015, Buder2025b}. We focus first on a representative three-night GALAH sample. This keeps the analysis compact and transparent while still sampling ordinary survey targets across the broad GALAH magnitude range and 2-degree field of view \citep{Martell2017}, rather than selecting stars a posteriori by stellar labels or residual behaviour. A full survey-wide residual model is left to future work. We also use external activity indicators to test whether residuals encode spectral variability that is not part of the single-star photospheric synthesis.

Using the GALAH data and samples introduced in Sec.~\ref{sec:data}, we address the broader question of what information remains in stellar-spectrum residuals through two complementary analyses. We first audit the significance, sign, recurrence, and plausible origins of the residual structure (Sec.~\ref{sec:analysis_audit}), and then use data-driven models to determine which fitted and unfitted properties can still be recovered from the residual spectra alone (Sec.~\ref{sec:analysis_cannon}). These analyses address four specific questions, discussed further in Sec.~\ref{sec:discussion}:

\begin{enumerate}[leftmargin=1.3em,labelwidth=0em]
    \item How significant, systematic, and recurrent are the remaining residuals relative to the observational uncertainties, and how do they separate into features that are over-predicted or under-predicted by the synthesis?
    \item Which residual features can be associated with plausible stellar, interstellar, sky, instrumental, reduction, or line data origins, and which of these provide actionable targets for improving the spectral modelling?
    \item To what extent do residual spectra retain information about properties already included in the synthesis fit, such as stellar parameters and elemental abundance patterns, as well as properties not explicitly fitted, such as interstellar extinction and chromospheric activity?
    \item Where within the spectrum is this information concentrated, and in particular, how much remains in wavelength regions excluded from the original GALAH DR4 optimisation?
\end{enumerate}

Residual modelling is used here as a diagnostic rather than a replacement for physical synthesis: the aim is to identify where improved line data, line profiles, physical models, reduction procedures, additional labels, or complementary data-driven corrections would recover information that the current model leaves behind.

\section{Data} \label{sec:data}

For each spectrum, we use the observed continuum-normalised flux, its uncertainty, the best-fitting synthetic flux, and reference labels or diagnostics. Section~\ref{sec:data_galah} describes these GALAH DR4 products, while Sec.~\ref{sec:data_samples} defines the parent three-night sample and the baseline, abundance, and activity subsamples.

\subsection{GALAH DR4 spectra and labels}
\label{sec:data_galah}

We use data products from GALAH DR4 \citep{Buder2025b}, based on spectra observed with the HERMES spectrograph \citep{Barden2010, Brzeski2011, Heijmans2012, Farrell2014, Sheinis2015} on the Anglo-Australian Telescope as part of the GALAH survey \citep{DeSilva2015}. HERMES records high-resolution optical spectra with a nominal resolution of 28000 in four non-contiguous wavelength channels, covering the blue ($4713$--$4903\,\text{\AA}$), green ($5648$--$5873\,\text{\AA}$), red ($6478$--$6737\,\text{\AA}$), and near-infrared ($7585$--$7887\,\text{\AA}$) arms for up to 400 fibres of the 2dF position system \citep{Lewis2002, Miszalski2006}. GALAH DR4 combines observations from the main GALAH survey with K2-HERMES, TESS-HERMES, cluster, halo, and other targeted HERMES programs. We return to the specific target selection of the representative sample in Section~\ref{sec:data_samples}.

\begin{figure}
    \centering
    \includegraphics[width=\linewidth]{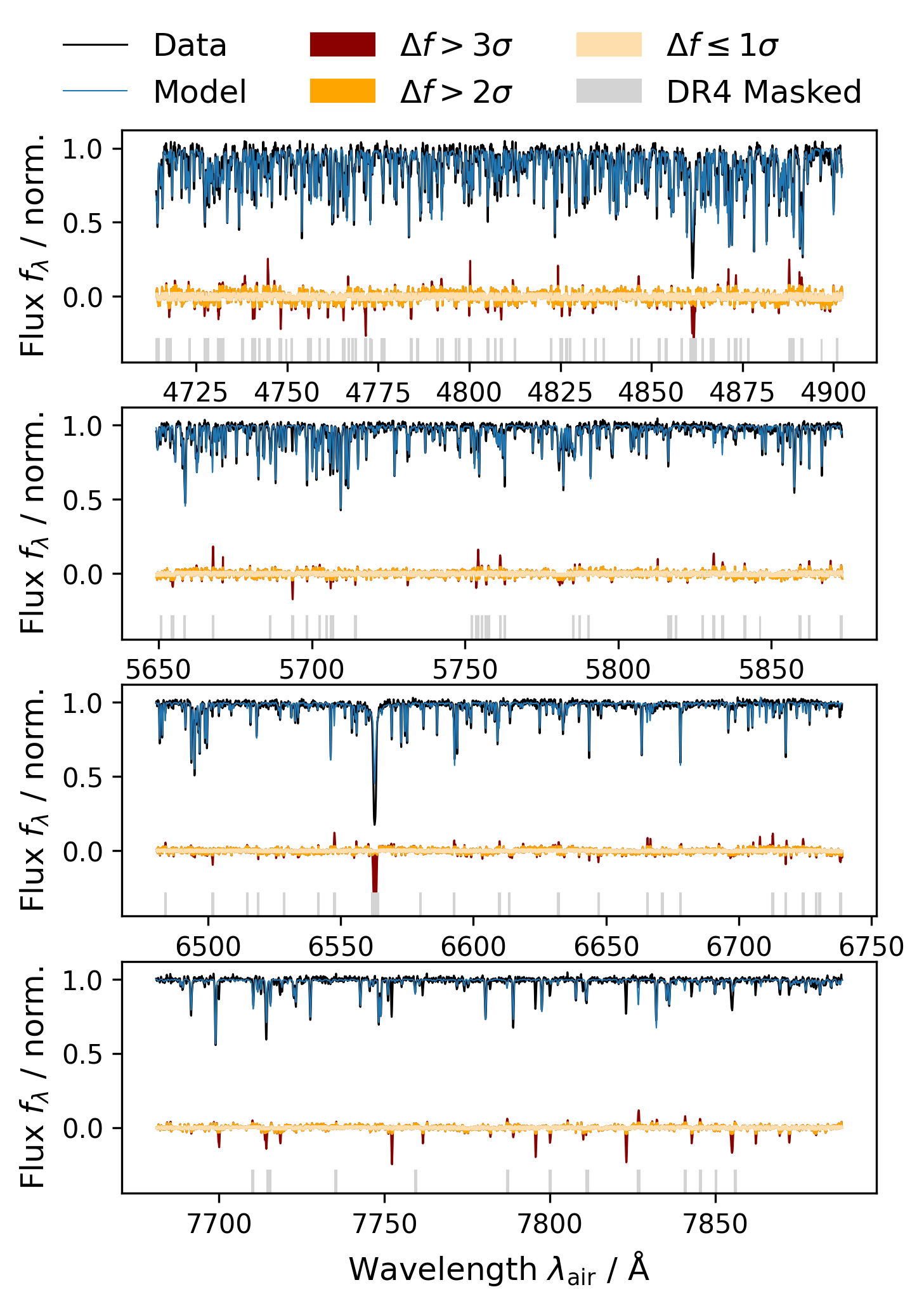}
    \caption{Example GALAH DR4 spectrum of the red giant 2MASS~J22084160+0607286 (\texttt{sobject\_id} 170710003201361). Black and blue lines show the observed continuum-normalised spectrum and the best-fitting DR4 synthesis, respectively, across the four HERMES CCD arms. Shaded regions show the residual flux, with darker shades indicating larger significance relative to the reported flux uncertainty. Grey bands mark wavelength regions excluded from the DR4 optimisation.}
    \label{fig:example_spectrum_170710003201361}
\end{figure}

The reduced 1D spectra are produced with the HERMES reduction pipeline of \citet{Kos2017}, updated for GALAH DR4. This pipeline performs the extraction, wavelength calibration, sky subtraction, telluric correction, barycentric correction, uncertainty estimation, and line-spread-function measurement for each spectrum. GALAH DR4 provides the observed flux, the associated flux uncertainty, and the best-fitting synthetic spectrum inferred by the DR4 analysis and used to normalise the observation, as exemplified in Fig.~\ref{fig:example_spectrum_170710003201361}, where we visualise residuals in three different intensities based on their significance with respect to the reported flux uncertainty.

In using these products, we retain the GALAH DR4 masking of the strongly telluric-contaminated region $7585$--$7680\,\text{\AA}$. A reliable treatment of this wavelength range requires dedicated telluric modelling alongside the stellar model, as demonstrated by \citet{Saydjari2023} and \citet{Casey2026}, whereas our aim is to assess residual structure in the DR4 analysis products rather than to introduce an additional telluric-modelling step.

The GALAH DR4 stellar labels were inferred from a hybrid spectrum-synthesis workflow. The synthetic spectra were computed with \textsc{Spectroscopy Made Easy} \citep[SME;][]{Valenti1996, Piskunov2017} and 1D MARCS model atmospheres \citep{Gustafsson2008}, using the GALAH line list developed from the compilation of \citet{Heiter2021}. To make the analysis feasible across the full GALAH wavelength range, GALAH DR4 trains neural-network interpolators on synthetic spectra in local regions of the $T_\mathrm{eff}$, $\log g$, and $[\mathrm{Fe/H}]$ parameter space. These interpolators enable the simultaneous optimisation of global stellar parameters and elemental abundances across most of the observed wavelength range, including broad molecular C$_2$ and CN absorption features. The optimisation was done using up to 94\% of the pixels, with several pixels being excluded by default after initial tests indicated persistently poor model--data agreement \citep[for more details see Sec.~4.3.2 of][]{Buder2025b}. The 6\% of pixels that were synthesised but not used for optimisation are particularly relevant here and are shown as grey regions in Fig.~\ref{fig:example_spectrum_170710003201361}.

The synthetic spectra include 1D non-LTE line formation for atomic lines where available. In GALAH DR4, this treatment is implemented through departure-coefficient grids from the BALDER/Multi3D framework and associated model atoms for H, Li, C, N, O, Na, Mg, Al, Si, K, Ca, Mn, Fe, and Ba \citep[e.g.][]{Amarsi2020, Amarsi2022}. Other atomic transitions and molecular features are treated in 1D LTE. This distinction is important for the interpretation of residuals: although GALAH DR4 models the broad C$_2$ and CN absorption that affected earlier analyses, these molecular features do not yet have the same non-LTE treatment as the listed atomic transitions.

We use the GALAH DR4 labels as reference quantities against which residual structure can be tested. The fitted stellar labels include the global stellar parameters $T_\mathrm{eff}$, $\log g$, $[\mathrm{Fe/H}]$, microturbulence velocity $v_\mathrm{mic}$, projected rotational velocity $v\sin i$, and radial velocity $v_\mathrm{rad}$ as well as elemental abundances for Li, C, N, O, Na, Mg, Al, Si, K, Ca, Sc, Ti, V, Cr, Mn, Fe, Co, Ni, Cu, Zn, Rb, Sr, Y, Zr, Mo, Ru, Ba, La, Ce, Nd, Sm, and Eu. We use the associated abundance flags to distinguish reliable detections from upper limits, non-detections, and measurements affected by optimisation or boundary issues.

GALAH DR4 reports spectra on their native pixel grids, with individual wavelength solutions for each observed spectrum. Because the data-driven models used in this work require all spectra to be represented on a shared wavelength grid, we interpolate the observed fluxes, best-fitting synthetic fluxes, and flux uncertainties onto a common wavelength grid for each HERMES arm. The construction of the common grid is described in Section~\ref{sec:data_samples}.

We also make use of auxiliary quantities reported by GALAH DR4. These include the extinction-related quantity $A_\mathrm{K_S}$ estimated via the Rayleigh-Jeans Color Excess method \citep{Majewski2011}, as reported by \citet{Buder2025b}. This property is not fitted by the spectral synthesis, but it provides a useful reference label for testing whether residual spectra contain information associated with line-of-sight dust and interstellar absorption. GALAH DR4 also reports Gaussian fits to residuals at the expected positions of interstellar K~{\sc i} and three diffuse interstellar bands \citep[see Sec.~6.1.3][]{Buder2025b}, as well as measurements of emission signals in the Balmer lines \citep[see Sec.~6.1.4][]{Buder2025b}. These are diagnostic products of the residual spectra and are not included in the stellar synthetic spectra used in the DR4 analysis.

\subsection{Parent sample and analysis samples}
\label{sec:data_samples}

We select 7627 unique stars observed in 22 nominal GALAH fields over 10--12 July 2017, with no repeat observations. The pointings comprise 
three bright fields ($9<V/\mathrm{mag}<12$), 
18 main fields ($12<V/\mathrm{mag}<14$), and 
one faint field ($9<V/\mathrm{mag}<14.3$). 
Because targets were randomly selected within the standard field magnitude limits rather than by stellar labels or residual behaviour, this provides a compact but representative pilot sample. We derive a \textit{baseline} sample and more restrictive \textit{abundance} and \textit{activity} subsamples below. The \textit{baseline} sample supports the general residual audit and global label experiments, the more restrictive \textit{abundance} sample increases sensitivity to chemical abundance correlations, and the \textit{activity} sample tests whether residuals retain information about an independently measured property that was not fitted by the GALAH synthesis.

For all analysis samples, we adopt a common residual representation on shared wavelength grids defined separately for each HERMES arm. We define each grid using the largest starting wavelength, the median wavelength dispersion, and the smallest final wavelength among the selected spectra. This gives common wavelength ranges of
\begin{align}
    4719.0299..&0.0460..4893.8808\,\text{\AA}, \notag \\
    5655.5298..&0.0547..5863.0182\,\text{\AA}, \notag \\
    6486.0104..&0.0631..6726.4572\,\text{\AA}\text{, and} \notag \\
    7689.1261..&0.0735..7874.3127\,\text{\AA} \notag
\end{align}
for the blue, green, red, and near-infrared arms, respectively. We linearly interpolate the observed flux, best-fitting synthetic flux, and flux uncertainty onto these grids before constructing the residual matrix.

We define the residual spectrum as the difference between the observed and best-fitting synthetic flux,
\begin{equation}
    \Delta f_\lambda = f_{\lambda\,\mathrm{obs}} - f_{\lambda\,\mathrm{model}}, \label{eq:def_residuals}
\end{equation}
and use the reported flux uncertainties $\sigma_{f_\lambda}$ to calculate a residual significance,
\begin{equation}
    s_\lambda = \frac{\vert \Delta f_\lambda \vert}{\sigma_{f_\lambda}}, \label{eq:def_significance}
\end{equation}
and to construct inverse-variance weights,
\begin{equation}
    w_\lambda = \sigma_{f_\lambda}^{-2}. \label{eq:def_ivar}
\end{equation}
After requiring finite interpolated fluxes and uncertainties across the common wavelength grid, the final matrix contains 7087 spectra and 13927 wavelength pixels. This residual matrix forms the input to the data-driven models analysed in the later sections.

\begin{table}
\centering
\caption{Overview of the wavelength pixels, samples, and labels used by the residual models. See Secs.~\ref{sec:data_samples} and \ref{sec:analysis_cannon} for details on samples/models.}
\label{tab:residual_models}
\begin{tabular}{lccccccccccc}
\hline\hline
Model & $N_\lambda$ & $N_\star$ & $T/g/\mathrm{Fe}$ & $\mathrm{[X/Fe]}$ & $A_\mathrm{K_S}$ & $\log R_\mathrm{HK}'$ \\
\hline
\textcolor{M1}{M1} & 13927 & 7087 & $\times$ & -- & -- & -- \\
\textcolor{M1}{M1*} & 13927 & 3826 & $\times$ & -- & -- & -- \\
\textcolor{M2}{M2} & 13927 & 3826 & $\times$ & $\times$ & -- & -- \\
\textcolor{M3}{M3} & 13927 & 7087 & $\times$ & -- & $\times$ & -- \\
\textcolor{M4}{M4} & 13927 &  226 & $\times$ & -- &  --  & $\times$ \\
\textcolor{M5}{M5} &   793 & 7087 & $\times$ & -- & -- & -- \\
\textcolor{M6}{M6} & 13134 & 7087 & $\times$ & -- & -- & -- \\
\hline\hline
\end{tabular} \newline
\footnotesize{Note: Each model uses a certain number of pixels ($N_\lambda$) for a total of $N_\star$ spectra with input labels $T_\mathrm{eff}$, $\log g$, and $\mathrm{[Fe/H]}$ (abbreviated as $T/g/\mathrm{Fe}$). $\mathrm{[X/Fe]}$ denotes the additional use of abundances of elements Na, Mg, Mn, and Ba. Crosses indicate labels included in each model. Model \textcolor{M5}{M5} uses only pixels excluded from the GALAH DR4 optimisation, while \textcolor{M6}{M6} uses only pixels included in it.}
\end{table}

We construct the residual models listed in Tab.~\ref{tab:residual_models} from different samples, labels, and wavelength subsets. Model~\textcolor{M1}{M1} provides the baseline test using $T_\mathrm{eff}$, $\log g$, and $[\mathrm{Fe/H}]$; subsequent models add abundance, extinction, or activity labels, or restrict the wavelength coverage. All models are formally introduced later in Sec.~\ref{sec:analysis_cannon}.

\subsubsection{Baseline Sample}
\label{sec:sample_baseline}

We require a converged DR4 solution ($\texttt{flag\_sp}<2048$), $\chi^2_\mathrm{sp}<2$, finite $A_\mathrm{K_S}$, and signal-to-noise ratios $\texttt{snr\_px\_ccd}>10$ in all four CCDs. We retain spectra with emission-line or binary flags because they test whether non-stellar or poorly modelled structure is recovered. These criteria leave 7087 unique spectra with median signal-to-noise ratios of 29, 42, 55, and 50 across the four CCDs.

\begin{figure*}
    \centering
    \includegraphics[width=\textwidth]{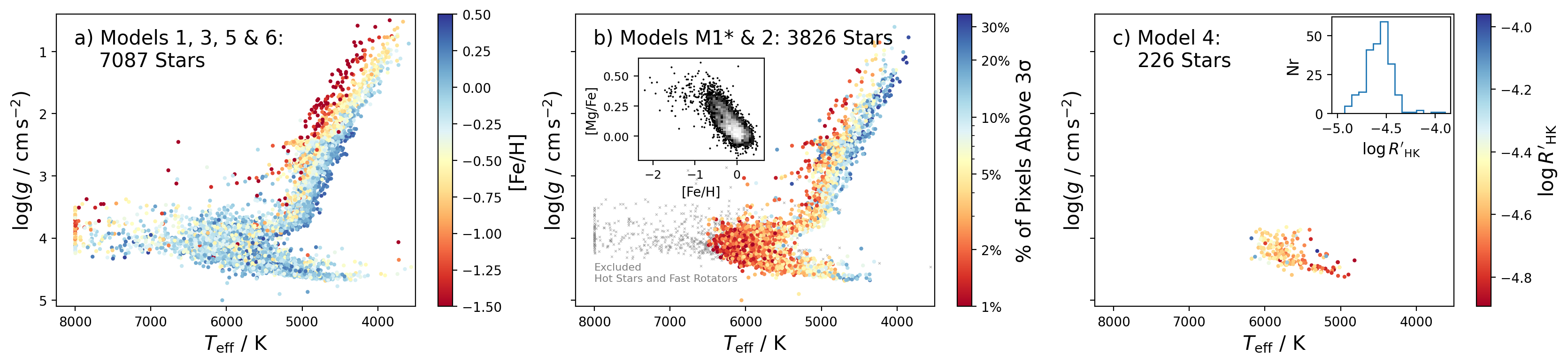}
    \caption{Stellar-label distributions of the three analysis samples in the $T_\mathrm{eff}$--$\log g$ plane. \textit{Left:} the 7087-star \textit{baseline} sample used for Models~\textcolor{M1}{M1}, \textcolor{M3}{M3}, \textcolor{M5}{M5}, and \textcolor{M6}{M6}, colour-coded by $[\mathrm{Fe/H}]$. \textit{Middle:} the 3826-star \textit{abundance} sample used for Models~\textcolor{M1}{M1*} and \textcolor{M2}{M2}, coloured by the percentage of pixels with residual significance above $3\sigma$; excluded hot stars and fast rotators are shown as grey crosses, and the inset shows the selected stars in the $[\mathrm{Mg/Fe}]$--$[\mathrm{Fe/H}]$ plane. \textit{Right:} the 226-star \textit{activity} sample used for Model~\textcolor{M4}{M4}, coloured by the activity indicator $\log R'_\mathrm{HK}$ from \citet{Zhang2024} with the $\log R'_\mathrm{HK}$ distribution in an inset.}
    \label{fig:labels_samples_and_models}
\end{figure*}

The distribution of this sample in the Kiel diagram is shown in Fig.~\ref{fig:labels_samples_and_models}a. The sample covers the main stellar types observed by GALAH, spanning approximately $3600 < T_\mathrm{eff}/\mathrm{K} < 8000$, $0.5 < \log(g/\mathrm{cm\,s^{-2}}) < 5.0$, and $-3.0 < \mathrm{[Fe/H]} < 0.75$ (but mainly $-0.5 < \mathrm{[Fe/H]} < 0.1$). The vast majority of stars are observed at low extinction, $A_{K_S} = 0.02_{-0.02}^{+0.09}\,\mathrm{mag}$ up to a maximum of $A_{K_S} \leq 0.37\,\mathrm{mag}$.

\subsubsection{Abundance Sample}
\label{sec:sample_abundance}

Because elements with similar nucleosynthetic origins exhibit correlated abundance patterns \citep[e.g.][]{Kobayashi2020}, we use Mg, Mn, Na, and Ba as representatives of core-collapse-supernova, Type-Ia-supernova, odd-$Z$, and asymptotic-giant-branch enrichment, respectively. We require unflagged abundances for all four elements (\texttt{flag\_x\_fe} = 0), $v\sin i<15\,\mathrm{km\,s^{-1}}$, and $T_\mathrm{eff}<6500\,\mathrm{K}$. The rotation and temperature cuts concentrate absorption into fewer pixels and increases sensitivity to weak residual correlations. These criteria leave 3826 spectra in the \textit{abundance} sample, shown in Fig.~\ref{fig:labels_samples_and_models}b; excluded hot stars and fast rotators are shown in grey. The fraction of pixels with residual significance above $3\sigma$ is generally largest among the coolest and most metal-rich stars. Visual inspection of cool stars with particularly large residual fractions shows that much of this additional structure is associated with imperfectly modelled molecular absorption, including CN, $\mathrm{C}_2$, and TiO features.

We do not include a separate rapid neutron-capture process label, because Eu is not modelled sufficiently reliably in GALAH DR4 \citep{Kane2026}, Sm is detected too infrequently, and adding Nd would reduce the sample from 3826 to 2663 spectra while providing a pattern similar to the selected neutron-capture element Ba.



\begin{figure*}
    \raggedleft
    \includegraphics[width=0.985\linewidth]{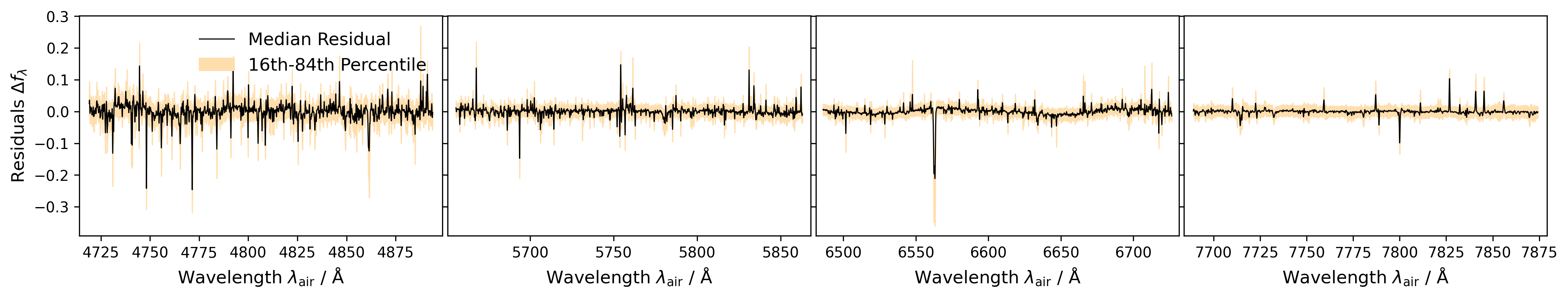}
    \includegraphics[width=0.979\linewidth]{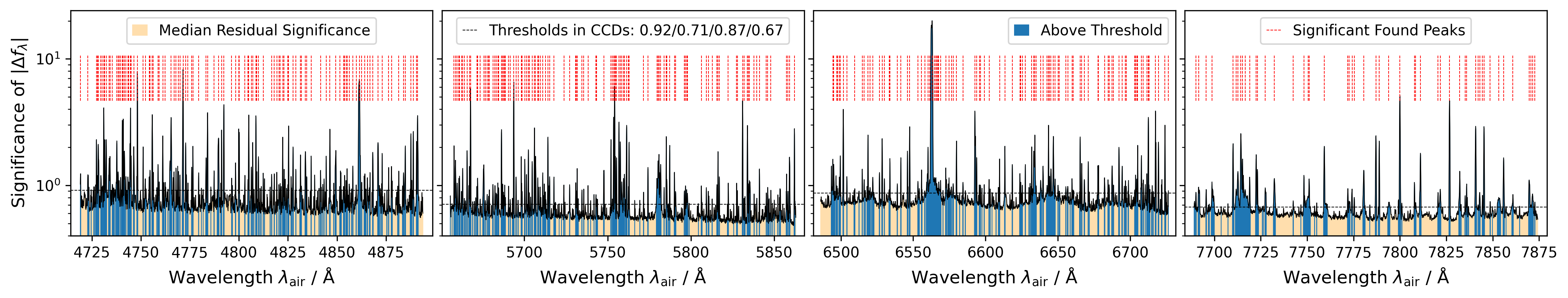}
    \includegraphics[width=\linewidth]{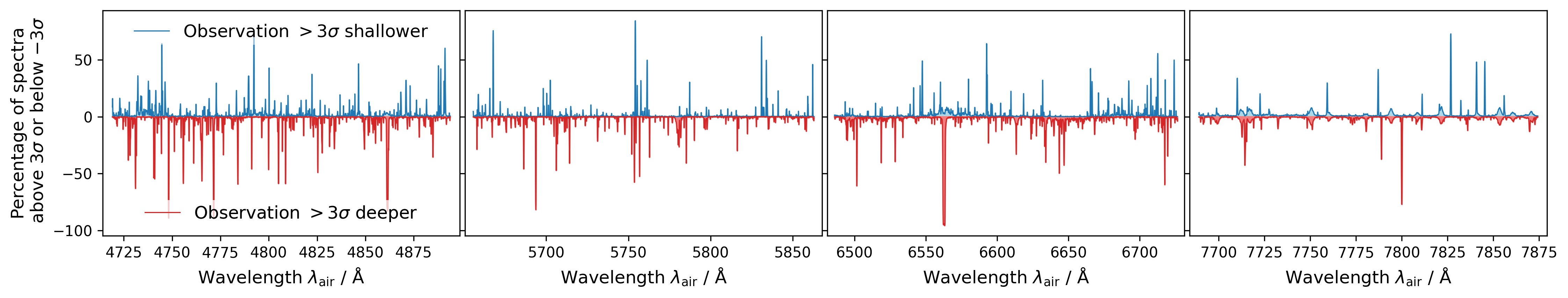}
    \caption{Residual statistics across the four GALAH wavelength bands. From top to bottom, the panels show the distribution of residual flux, the typical residual significance, and the fraction of the 7087 spectra for which the observed flux is significantly shallower (blue) or deeper (red) than the synthesis at more than $3\sigma$. Columns correspond to the four HERMES CCD arms. An enlarged version of the middle panels can be found in our \href{https://github.com/svenbuder/residual_cannon/blob/main/figures/significant\_residual\_peaks\_m1\_enlarged.png}{repository}.}
    \label{fig:residual_statistics}
\end{figure*}

\subsubsection{Activity Sample}
\label{sec:sample_activity}

To test whether residual spectra retain chromospheric information, we use the photosphere-corrected activity index $\log R'_\mathrm{HK}$, based on excess emission in the Ca~{\sc ii} H and K cores \citep[e.g.][]{Wilson1978,Noyes1984}. Although these lines lie outside the GALAH wavelength range, the LAMOST catalogue of \citet{Zhang2024} overlaps with 34\,672 GALAH DR4 entries\footnote{Overlaps with dedicated activity studies only yielded 83 stars from \citet{BoroSaikia2018}, 51 from \citet{Brown2022}, 40 from \citet{GomesDaSilva2021}, and one from \citet{LorenzoOliveira2018}.} and with 226 stars in our \textit{baseline} sample. Its G- and K-type dwarfs span $4800<T_\mathrm{eff}/\mathrm{K}<6250$, $\log(g/\mathrm{cm\,s^{-2}})>3.8$, and $-5.0<\log R'_\mathrm{HK}<-3.9$ (see Fig.~\ref{fig:labels_samples_and_models}c), covering the range in which \citet{Spina2020} found activity-dependent equivalent widths and abundances. To limit the influence of extreme positive sky-subtraction residuals, we cap continuum-normalised fluxes at 1.2 and assign the original flux value as their uncertainty before calculating $\Delta f$.

\section{A systematic audit of spectrum residuals}
\label{sec:analysis_audit}

We first test whether the residuals are consistent with random noise or contain coherent structure across the \textit{baseline} sample. We quantify their significance and sign as functions of wavelength before constructing a reproducible catalogue of their widths, recurrence, and plausible origins.

\subsection{Where do significant residuals occur?}
\label{sec:analysis_significance}

Negative residuals, $\Delta f_\lambda<0$, indicate that the observed absorption is deeper than the synthesis, whereas positive residuals indicate shallower absorption. This distinction helps separate missing or underestimated opacity from excessive absorption, line-core filling, normalisation, or reduction effects.

As a first global check, we compare the distribution of all residual significances in the baseline sample to the expectation for Gaussian noise. For normally distributed residuals with correctly estimated uncertainties, 68.3\% and $99.7\,\mathrm{\%}$ of residuals should lie within $1\sigma$ and $3\sigma$, respectively. Across the 99 million residual pixels in the baseline sample, only $63.4\,\mathrm{\%}$ and $95.3\,\mathrm{\%}$ lie within these limits. This corresponds to 4.6 million residual pixels outside $3\sigma$, compared to only 0.3 million expected for Gaussian noise. The residual field is therefore not fully described by the quoted observational uncertainties and random noise alone.

We next ask where these residuals occur in wavelength. Fig.~\ref{fig:residual_statistics} summarises three complementary diagnostics across the four GALAH wavelength arms. The top panels show the median signed residual across the 7087 spectra at each wavelength pixel, together with the 16th and 84th percentiles. Most pixels have median residuals close to zero, but several regions show coherent offsets in flux space. Among the 14 strongest negative median residuals, with $\Delta f_\lambda < -0.1$, 11 occur in the blue arm, including the strongest features near 4748, 4771, and $4861\,\text{\AA}$ ($\mathrm{H}\upbeta$). The remaining strong negative features occur near $5694\,\text{\AA}$, $\mathrm{H}\upalpha$ at 6563\,\AA, and the region near $7800\,\text{\AA}$. Strong positive median residuals are slightly less frequent; the eight clearest positive features ($\Delta f_\lambda > 0.1$) occur near 4745, 4792, 4888, 4891, 5668, 5754, 5831, and $7827\,\text{\AA}$. In addition to these narrow features, we identify broader low-amplitude structures, including a wave-like residual pattern in CCD3 and broader residual behaviour in CCD1. These broad structures are less obvious in absolute flux, but may still be significant relative to the uncertainties.

The middle panels of Fig.~\ref{fig:residual_statistics} show the median residual significance, that is, the median across the 7087 stars of $s_\lambda = \vert \Delta f_\lambda \vert / \sigma_{f_\lambda}$ as a function of wavelength. This view highlights wavelength regions where the residuals are systematically large compared to the flux uncertainties, even when the median signed residual is modest. The cumulative distribution of per-wavelength median significances confirms that most wavelength pixels are well behaved: $87.8\,\mathrm{\%}$ of pixels have a median residual significance below $1\sigma$, and $97.3\,\mathrm{\%}$ below $2\sigma$. The remaining tail, however, is precisely the important part for this audit. It identifies wavelength regions where the mismatch between observations and synthesis is not driven by a few individual outliers, but recurs across a substantial fraction of the baseline sample.

The typical level of residual significance also differs between the GALAH wavelength arms. CCD1 and CCD3 show slightly higher median residual significances than CCD2 and CCD4. This could indicate that the spectral synthesis performs differently across the arms, that the flux uncertainties are calibrated differently, or that some arms contain more difficult stellar, telluric, or reduction features. We therefore do not define significant residual regions using a single absolute threshold across all arms. Instead, in the next subsection, we identify outlying residual features relative to the median-significance distribution within each arm. Specifically, we use the 75th percentile of the median residual significance in each CCD as an arm-dependent reference level, corresponding to median significances of 0.92, 0.71, 0.87, and 0.67 for CCDs 1-4, respectively. These levels are shown as dashed horizontal lines in the middle panels of Fig.~\ref{fig:residual_statistics}. Pixels above these thresholds (highlighted as blue regions) reveal both narrow residual peaks and broader residual structures, including the $\mathrm{H}\upalpha$ region and broad structure around $7715\,\text{\AA}$.

Finally, we quantify how often residuals occur with a preferred sign. For each wavelength pixel, we calculate the fraction of the 7087 spectra in which the observed flux is significantly shallower or deeper than the synthesis. The bottom panels of Fig.~\ref{fig:residual_statistics} show these signed recurrence fractions for the $3\sigma$ threshold. Most wavelength pixels have low recurrence fractions in both directions, but several features are significant in the same direction for a large fraction of the sample. We find nine wavelength regions where the observations are significantly shallower than the synthesis for more than $50\,\mathrm{\%}$ of spectra: 4745, 4792, 4891, 5668, 5754, 5831, 6593, 6713, and $7827\,\text{\AA}$. Conversely, we find 18 wavelength regions where the observations are significantly deeper than the synthesis for more than $50\,\mathrm{\%}$ of spectra: 4731, 4740, 4741, 4748, 4756, 4765, 4771, 4784, 4805, 4809, 4861, 5694, 5754, 5757, 6502, 6563, 6718, and $7800\,\text{\AA}$. The Balmer lines are dominated by residuals in which the observed spectra are deeper than the synthesis. Other regions, especially in CCD4, show broader residual structure with lower but noticeable recurrence in both directions, suggesting residual behaviour that may vary with stellar type, observing conditions, or reduction quality. The wave-like pattern in CCD3 is also visible in the signed recurrence diagnostic, although much of it remains below the $3\sigma$ threshold and is more apparent in the corresponding $1\sigma$ and $2\sigma$ panels shown in Fig.~\ref{fig:percentage_residuals_1_and_2_sigma}.

\begin{figure*}
\centering
\includegraphics[width=0.68\linewidth]{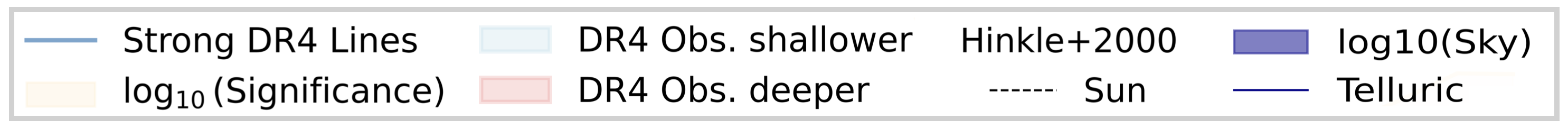}
\includegraphics[width=0.33\linewidth]{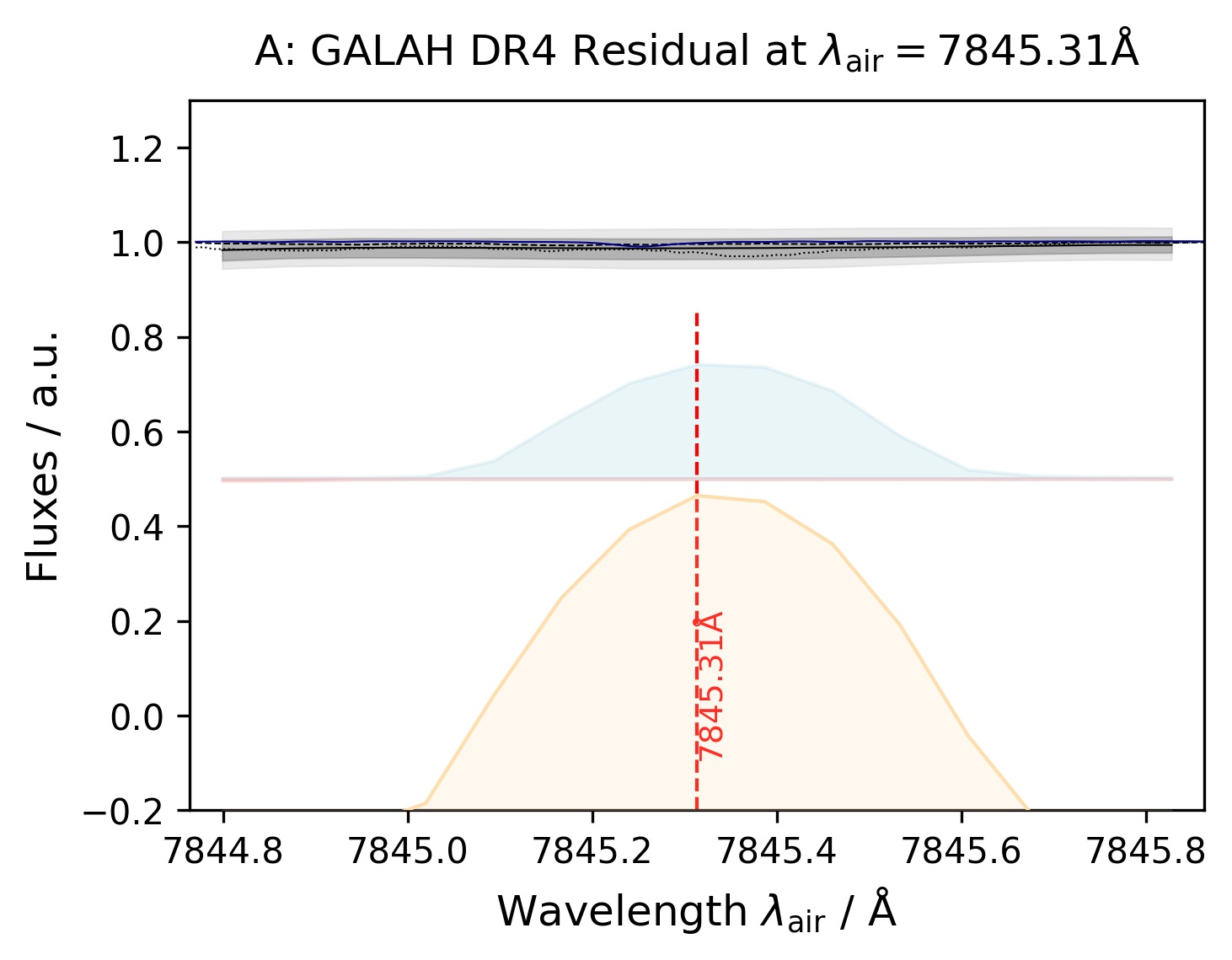}
\includegraphics[width=0.33\linewidth]{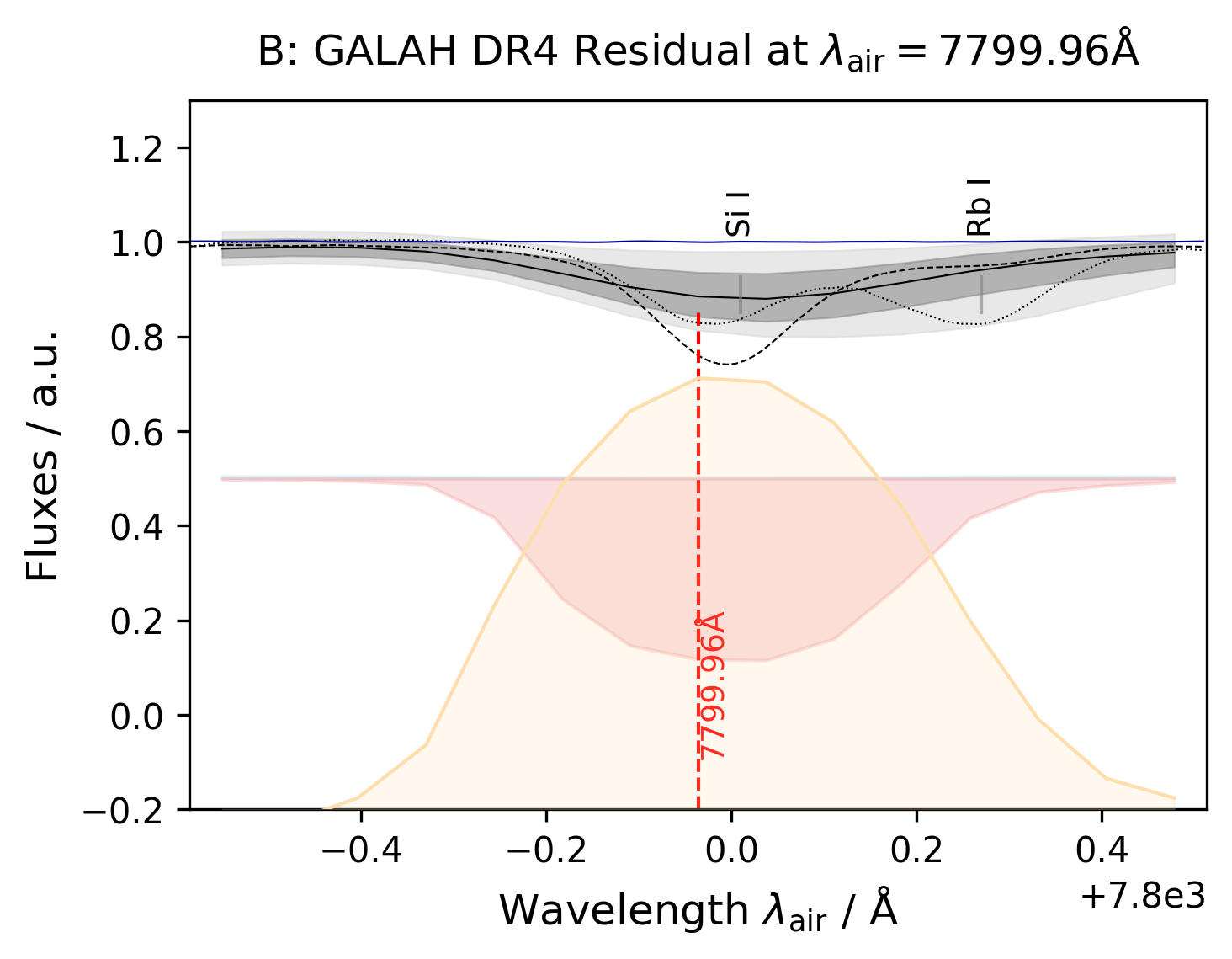}
\includegraphics[width=0.33\linewidth]{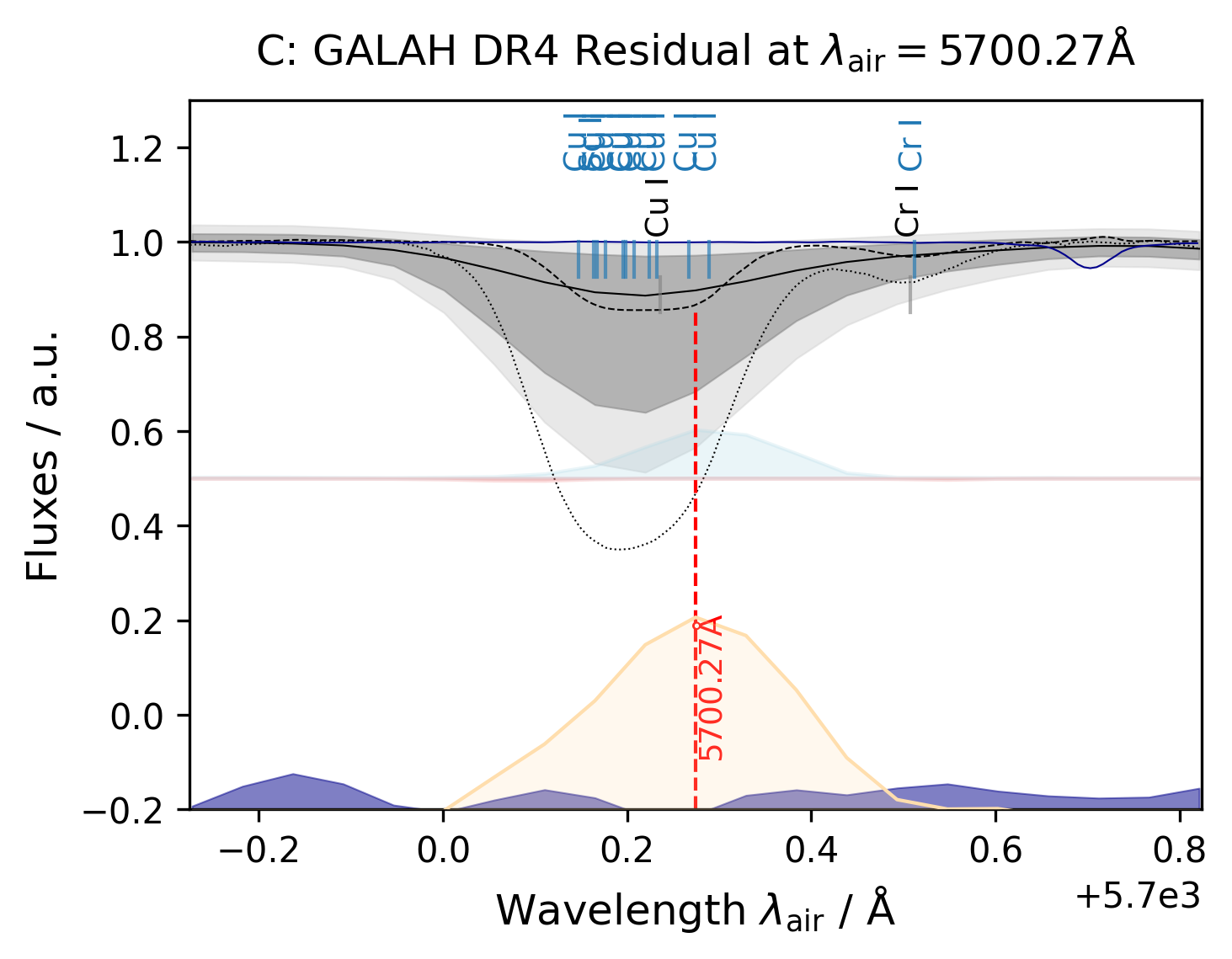}
\includegraphics[width=0.33\linewidth]{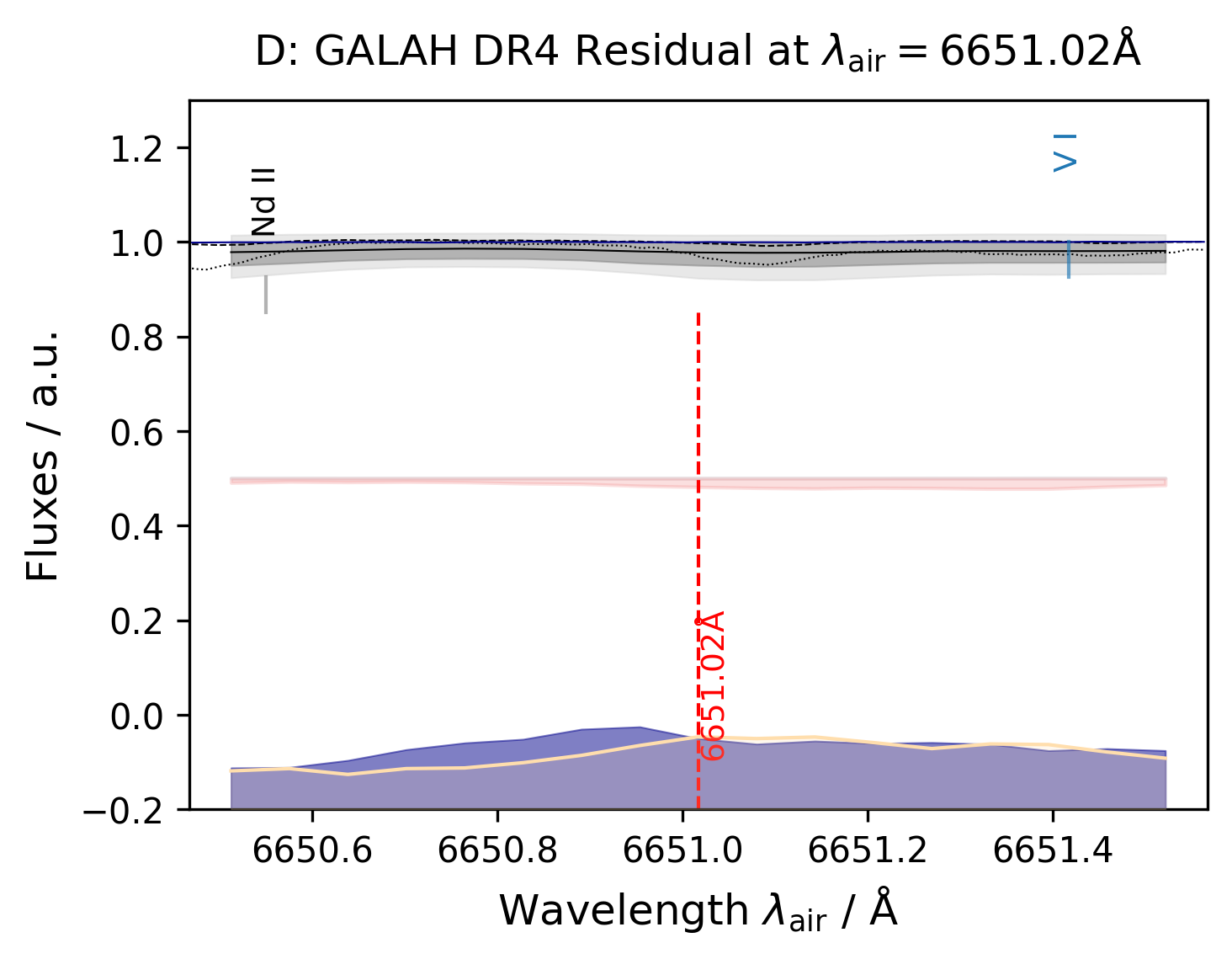}
\includegraphics[width=0.33\linewidth]{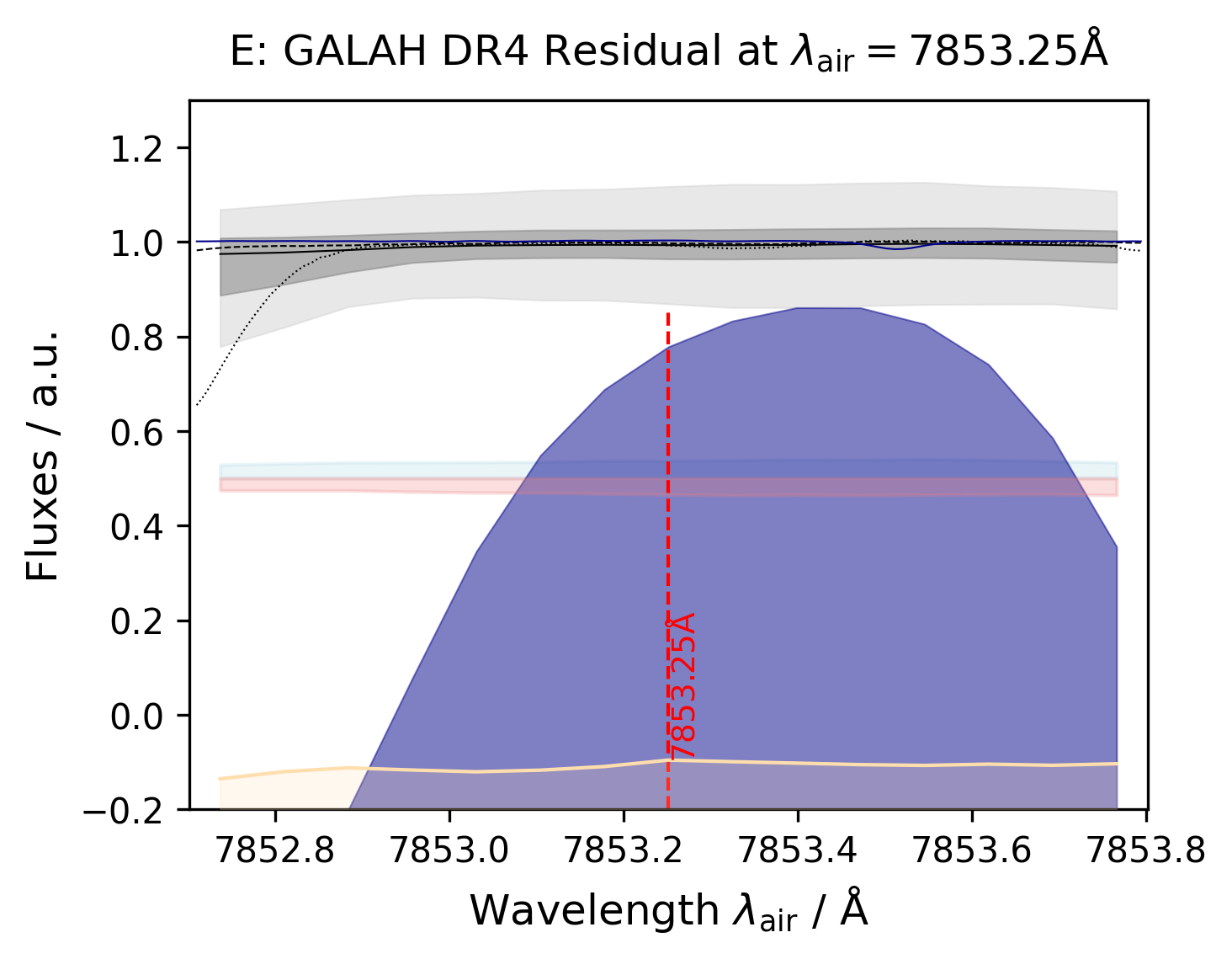}
\includegraphics[width=0.33\linewidth]{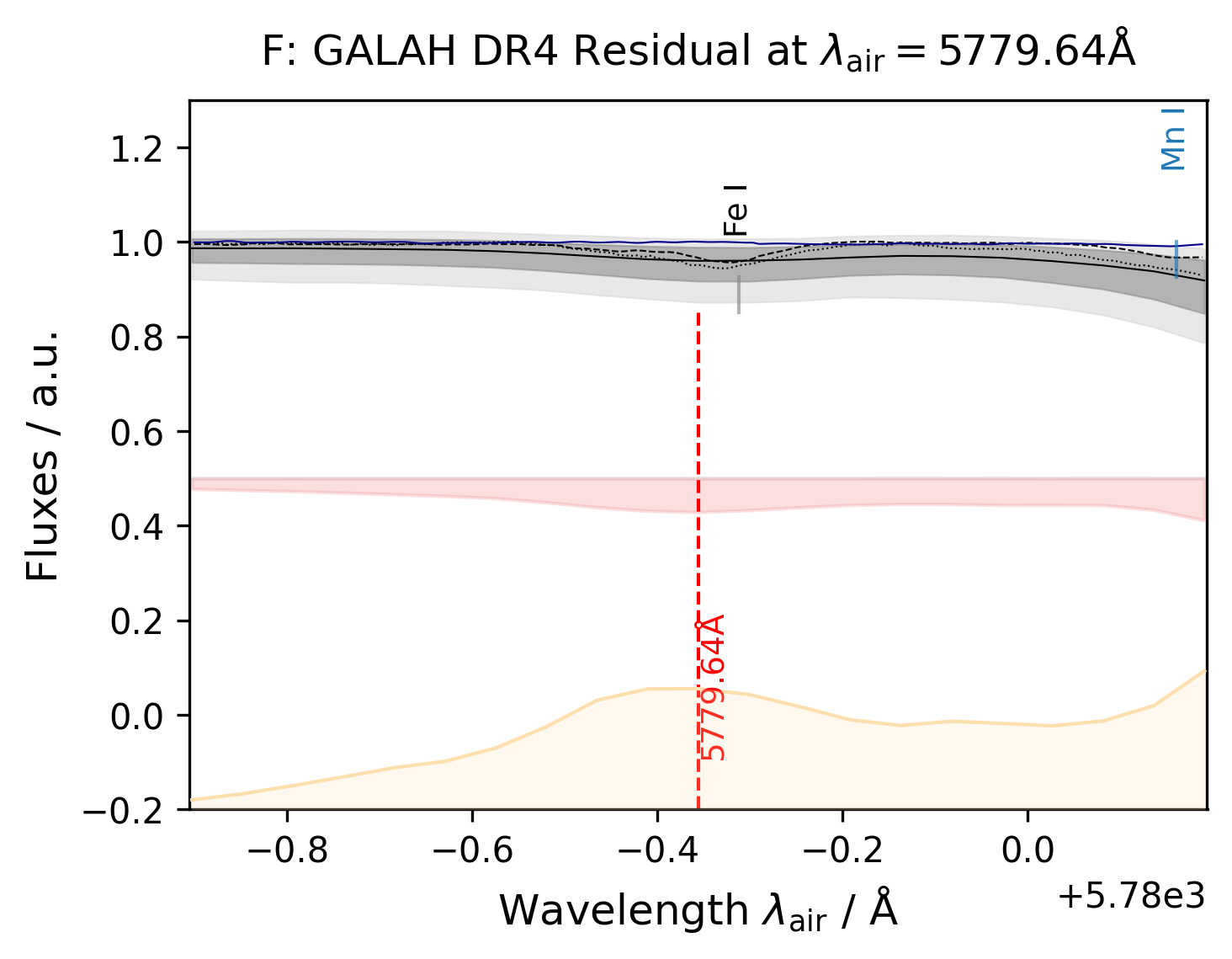}
\includegraphics[width=0.33\linewidth]{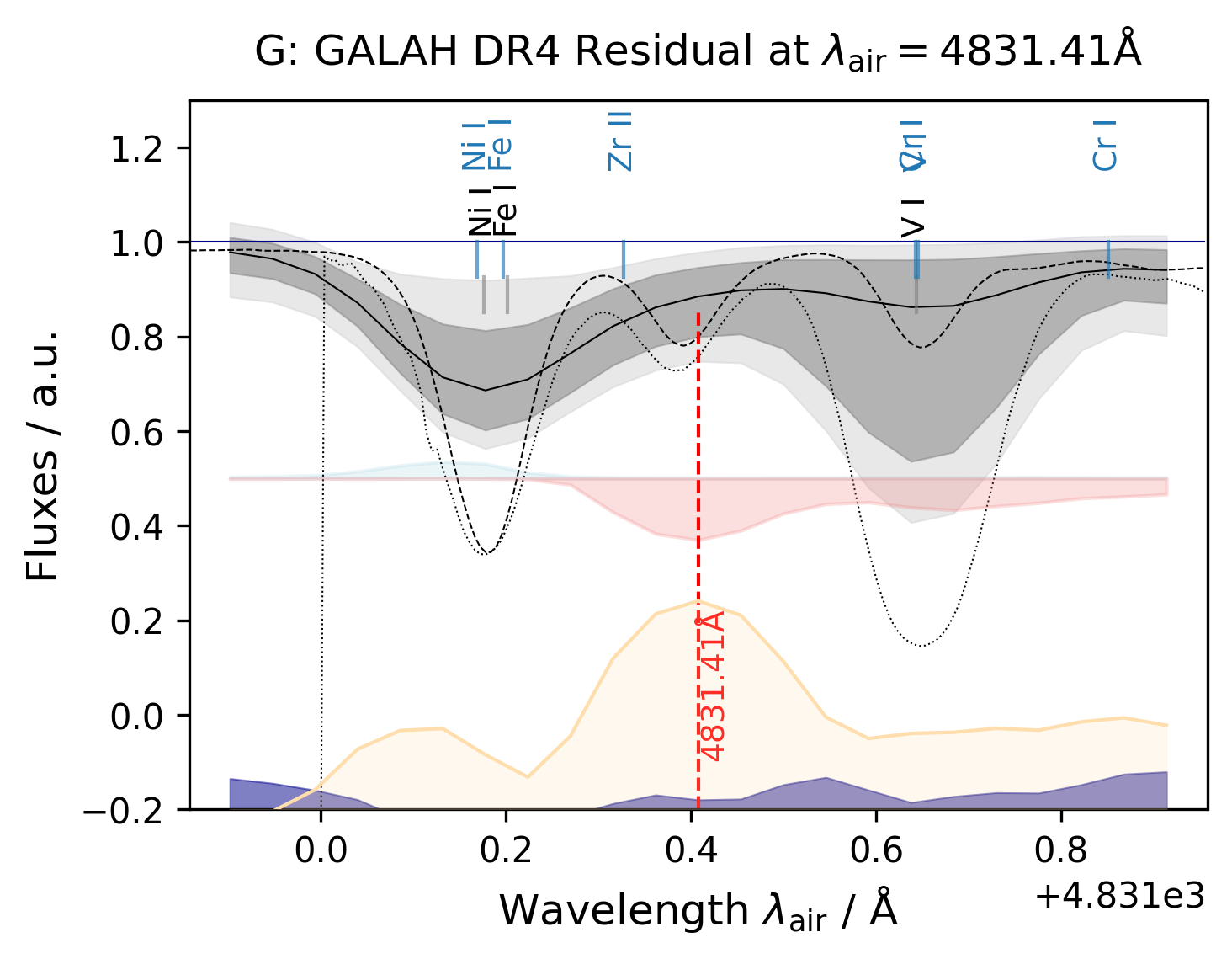}
\includegraphics[width=0.33\linewidth]{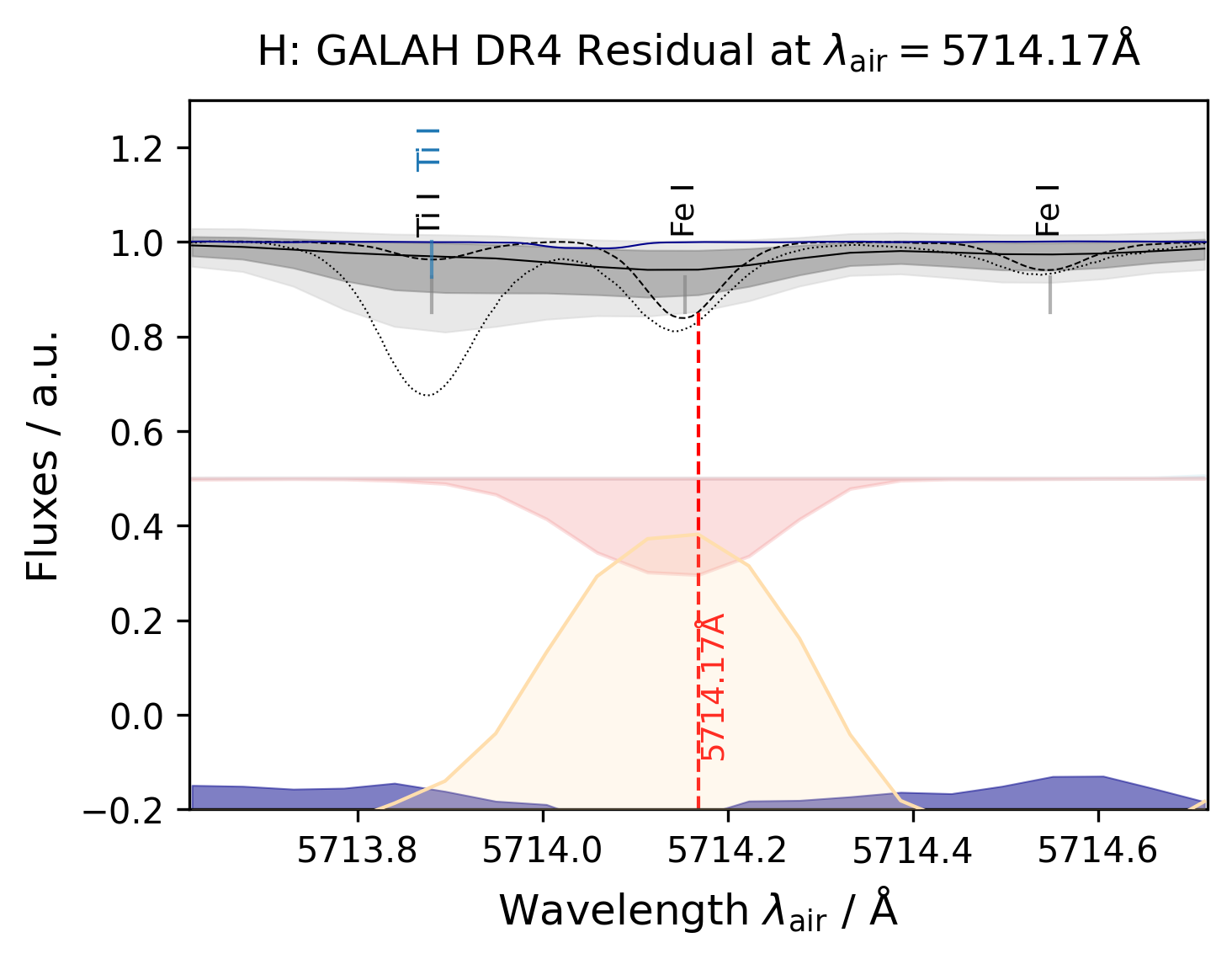}
\includegraphics[width=0.33\linewidth]{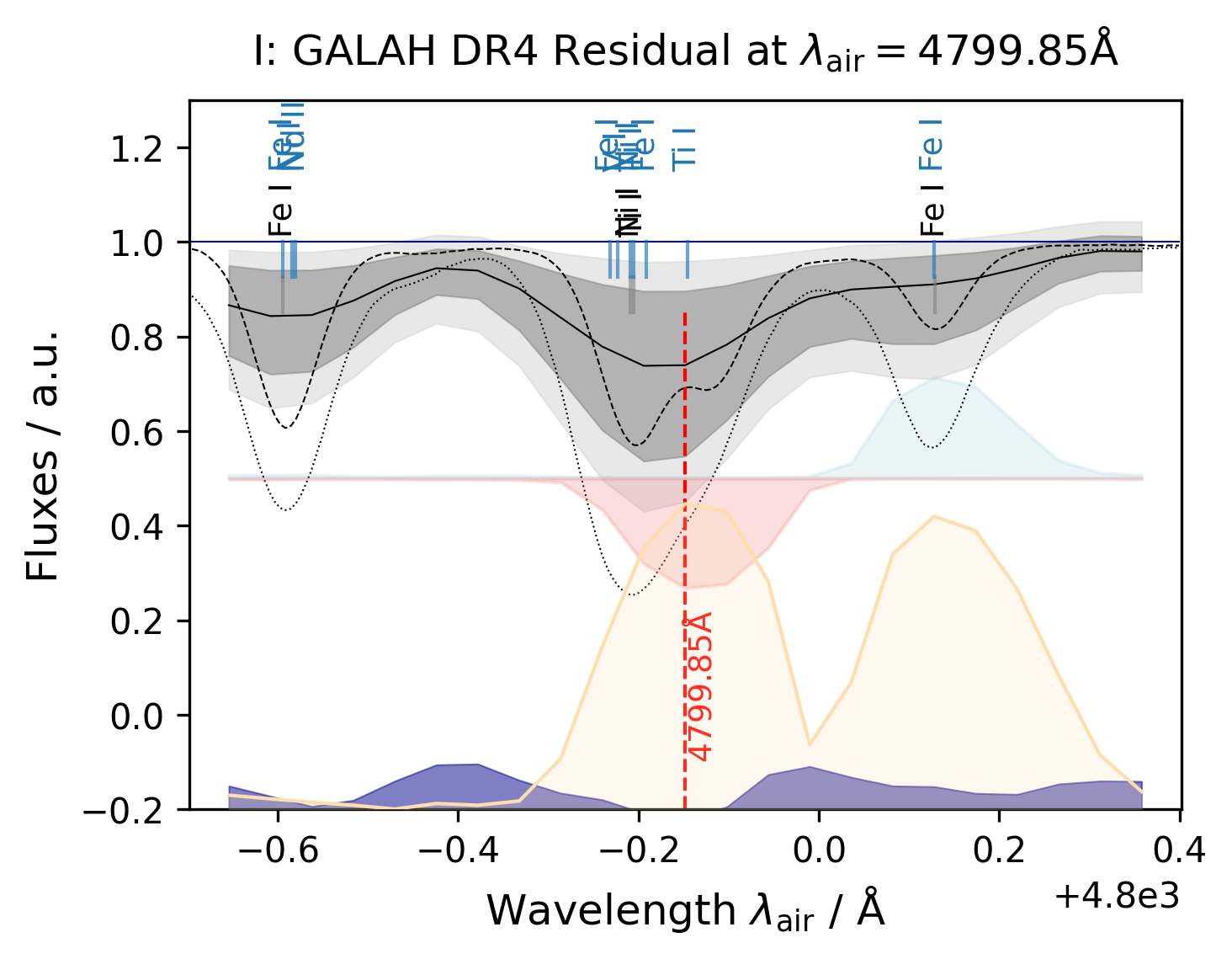}
\caption{Representative significant residual regions and their adopted classifications. Panels A--I illustrate, respectively, candidate \textit{$\log gf$ too high}, \textit{$\log gf$ too low}, \textit{Line profile}, \textit{Insignificant}, \textit{Skyline}, \textit{Interstellar}, \textit{Unknown line}, \textit{Known missing}, and \textit{Unclear} classifications. Each panel combines the GALAH residual statistics with relevant external diagnostics, including the Solar and Arcturus atlases, GALAH line-list entries, telluric absorption, and sky-emission measurements.}
\label{fig:example_significant_residual_regions}
\end{figure*}

\subsection{A catalogue of residual peaks and their plausible causes}
\label{sec:analysis_peak_catalogue}

We convert the wavelength-localised, recurrent residual structure into a reproducible catalogue of candidate regions. The catalogue supports inspection, classification, and future model improvement rather than providing a unique mathematical or physical decomposition.

We identify residual peaks from the median residual-significance spectrum of the \textit{baseline} sample. We use the median rather than the maximum significance so that the selection is driven by residuals recurring across many stars rather than by individual outliers. The search is performed independently in each GALAH wavelength arm using the \textsc{find\_peaks} function of \textsc{scipy} \citep{Scipy}. As motivated in Sec.~\ref{sec:analysis_significance}, we adopt the 75th percentile of the median residual-significance distribution in each arm as an arm-dependent reference level. We additionally require a minimum peak width of 3 pixels and a minimum separation of 6 pixels. These criteria suppress isolated single-pixel artefacts and reduce the subdivision of broad residual structures into multiple closely spaced detections. The resulting peak positions are shown as red dashed lines in the middle panels of Fig.~\ref{fig:residual_statistics}.

The adopted criteria identify 112%
, 117%
, 100%
, and 51%
 regions in CCDs 1--4, respectively, for a total of 380. Their exact number necessarily depends on the threshold, width, and separation parameters.

For each region, we create an inspection figure (with examples shown in Fig.~\ref{fig:example_significant_residual_regions}) combines the median logarithmic significance with the fractions of spectra showing significantly deeper or shallower observed flux. Mixed signs can indicate label-dependent behaviour or varying observational conditions, while broad regions are treated cautiously because observed-frame telluric, sky, and reduction features are smeared in the stellar rest frame.

We compare each region with the Solar and Arcturus atlases of \citet{Hinkle2000}, including their stellar-line identifications and telluric measurements. We also compare against the GALAH DR4 line list, selecting atomic entries with estimated central depths of $\texttt{DEPTH}>0.1$ for identifying possible contributors and $\texttt{DEPTH}>0.85$ for visualisation \citep[see also Sec.~4.2 of][]{Heiter2021}. This does not constitute a complete list of all possible blends, but includes the transitions expected to be most visible in high-quality Solar and Arcturus spectra. We additionally compare broad or ambiguous regions against the catalogue of diffuse interstellar bands by \citet{Vogrincic2023} and against sky-emission calibration observations associated with the observations shown in Fig.~\ref{fig:example_spectrum_170710003201361}.

\begin{table*}
\caption{Example entries from the significant-residual catalogue, including residual signifance $s_\lambda$ (see Eq.~\ref{eq:def_significance}) and sky flux $f_\mathrm{sky}$, selected to illustrate the nine classifications shown in Fig.~\ref{fig:example_significant_residual_regions}. The complete machine-readable catalogue is available in the public repository.}
\label{tab:tabular_peak_information}
\begin{tabular}{ccccccccccc}
\hline
Peak & $\lambda_\mathrm{air}$ & $s_\lambda$ & Width  & Classification & Shallower & Hinkle+(2000) & Nearest DR4 & $\log gf$ & Sky              & Comments \\
     & $\text{\AA}$           &             & pixels &                & Deeper    & Species       & Species \& \textsc{DEPTH} &  & $f_\mathrm{sky}$ &          \\
\hline
371 & 7845.31 & 2.9 & 6 & loggf too high & +48\% & None & Ca I 0.75 & -0.08 &  & Fig.~\ref{fig:example_significant_residual_regions}A \\
  &         &     &  &             & -0\% &  &  (7845.34+0.03Å) & (K07) &  \\
357 & 7799.96 & 5.1 & 6 & loggf too low & +0\% & Si I & Si I 0.30 & -2.21 &  & Fig.~\ref{fig:example_significant_residual_regions}B \\
  &         &     &  &             & -77\% & (7800.01+0.05Å) &  (7800.00+0.03Å) & (K07) &  \\
148 & 5700.27 & 1.6 & 6 & Line profile & +21\% & Cu I & Cu I 0.92 & -2.69 &  & Fig.~\ref{fig:example_significant_residual_regions}C \\
  &         &     &  &             & -0\% & (5700.24+0.04Å) &  (5700.27+0.01Å) & (KR|1989ZPhyD..11..287C) &  \\
296 & 6651.02 & 0.9 & 4 & Insignificant & +0\% & None & None &  &  & Fig.~\ref{fig:example_significant_residual_regions}D \\
  &         &     &  &             & -4\% &  &   &  &  &  \\
373 & 7853.25 & 0.8 & 25 & Skyline & +7\% & None & None &  & 11.5 & Fig.~\ref{fig:example_significant_residual_regions}E \\
  &         &     &  &             & -7\% &  &   &  &  &  \\
190 & 5779.64 & 1.1 & 4 & Interstellar & +0\% & Fe I & Fe II 0.22 & -6.34 &  & Fig.~\ref{fig:example_significant_residual_regions}F \\
  &         &     &  &             & -14\% & (5779.69+0.04Å) &  (5779.66+0.01Å) & (K13) &  \\
78 & 4831.41 & 1.7 & 6 & Unknown line & +0\% & None & None &  &  & Fig.~\ref{fig:example_significant_residual_regions}G \\
  &         &     &  &             & -26\% &  &   &  &  &  \\
160 & 5714.17 & 2.4 & 6 & Known missing & +0\% & Fe I & None &  &  & Fig.~\ref{fig:example_significant_residual_regions}H \\
  &         &     &  &             & -41\% & (5714.15+0.01Å) &   &  &  &  \\
56 & 4799.85 & 2.8 & 4 & Unclear & +0\% & None & Ti I 0.91 & -0.92 &  & Fig.~\ref{fig:example_significant_residual_regions}I \\
  &         &     &  &             & -46\% &  &  (4799.85+0.00Å) & (K10) &  \\
\dots & \dots & \dots & \dots & \dots & \dots & \dots & \dots & \dots & \dots & \dots \\
\hline
\end{tabular}
\footnotesize{$\log gf$ references: K14: \citet{K14}, 
K06: \citet{K06}, 
CB: \citet{CB}, 
BHN: \citet{BHN}, 
KP: \citet{KP}, 
K13: \citet{K13}, 
MFW: \citet{MFW}, 
GESMCHF: \citet{GESMCHF}, 
2014ApJS..215...20L: \citet{2014ApJS..215...20L}, 
1983MNRAS.204..883B,1989A\&A...208..157G: \citet{1983MNRAS.204..883B,1989A&A...208..157G}, 
GESB86: \citet{GESB86}, 
1998PhRvA..57.1652Y: \citet{1998PhRvA..57.1652Y}, 
GESHRL14: \citet{GESHRL14}, 
WSL: \citet{WSL}, 
KL-astro: GALAH manually adjusted, 
HLSC: \citet{HLSC}, 
DLSC: \citet{DLSC}, 
2009A\&A...497..611M: \citet{2009A&A...497..611M}, 
BWL: \citet{BWL}, 
1990JQSRT..43..207C: \citet{1990JQSRT..43..207C}, 
PQWB: \citet{PQWB}, 
WBb: \citet{WBb}, 
NIST10: \citet{NIST10}, 
2013ApJS..205...11L: \citet{2013ApJS..205...11L}, 
2009JPCRD..38..565W: \citet{2009JPCRD..38..565W}, 
2015ApJS..220...13L: \citet{2015ApJS..220...13L}, 
MC: \citet{MC}, 
RU: \citet{RU}, 
BK: \citet{BK}, 
MRW: \citet{MRW}, 
CBcor: \citet{CBcor}, 
GESB82c+BWL: \citet{GESB82c,BWL}, 
LSCI: \citet{LSCI}, 
2014MNRAS.441.3127R: \citet{2014MNRAS.441.3127R}, 
FMW: \citet{FMW}, 
LD-HS: \citet{LD-HS}, 
LGWSC: \citet{LGWSC}, 
K07: \citet{K07}, 
K09: \citet{K09}, 
SDL: \citet{SDL}, 
NIST: \citet{NIST}, 
MULT: \citet{MULT}, 
K10: \citet{K10}, 
1985A\&A...153..109W: \citet{1985A&A...153..109W}, 
K08: \citet{K08}, 
GC: \citet{GC}, 
BK+BWL: \citet{BK,BWL}, 
GESB86+BWL: \citet{GESB86,BWL}, 
BLQS: \citet{BLQS}, 
GESB82c: \citet{GESB82c}, 
GKOb: \citet{GKOb}, 
LD: \citet{LD}, 
WLN: \citet{WLN}, 
KR,1989ZPhyD..11..287C: \citet{KR,1989ZPhyD..11..287C}, 
2014ApJS..211...20W: \citet{2014ApJS..211...20W}, 
MA-astro: GALAH manually adjusted}

\end{table*}

We assigned each region a preliminary classification as an ordered set of explicit if-else rules with fixed, manually selected thresholds based on wavelength, width, recurrence, residual direction, and proximity to the external diagnostics described above. The thresholds were chosen to encode the physically motivated criteria described below rather than fitted to the final classifications. Broad features coincident with established diffuse interstellar bands or interstellar K~{\sc i} were classified as \textit{Interstellar}, while features coincident with sky emission were classified as \textit{Skyline}. Regions showing significant residuals in fewer than $5\,\mathrm{\%}$ of the spectra were labelled \textit{Insignificant}.

For the remaining regions, we searched for stellar transitions within one detector pixel of the residual peak. Where the \citet{Hinkle2000} atlas and the GALAH DR4 line list identified the same transition, residuals that were predominantly shallower or deeper than the synthesis were classified as candidate \textit{$\log gf$ too high} or \textit{$\log gf$ too low}, respectively. Features present in the atlas but absent from the GALAH DR4 line list were classified as \textit{Known missing}, while features without a counterpart in either source were classified as \textit{Unknown line}. Regions with asymmetric, split, or otherwise structured profiles that could not be explained by a simple change in line strength were assigned to \textit{Line profile}. The \textit{Unclear} category was retained for blended or ambiguous cases for which no single explanation was sufficiently supported.

We subsequently inspected all 380 regions visually. The preliminary classification was confirmed for 312 regions ($82\,\mathrm{\%}$), while 68 classifications ($18\,\mathrm{\%}$) were revised. Most revisions involved blended or conflicting line identifications, Balmer-line or hyperfine-structure residuals, the broad undulating pattern in CCD3, ambiguous skyline coincidences, or cases in which the residual morphology indicated a (profile) mismatch rather than a simple line-strength error.

The final catalogue is dominated by candidate stellar line-strength mismatches. As shown in the legend of Fig.~\ref{fig:example_significant_residual_regions}, we classify 227 regions as \textit{$\log gf$ too high} (130 regions) or \textit{$\log gf$ too low} (94 regions), with a further 13 regions assigned to \textit{Line profile}. Together, these 236 regions account for $62\,\mathrm{\%}$ of the catalogue. Another 46 regions correspond to absent or unidentified absorption features, comprising 8 \textit{Known missing} and 38 \textit{Unknown line} classifications. We identify 15 \textit{Skyline} and 12 \textit{Interstellar} regions, while 19 cases remain \textit{Unclear}. The remaining 51 regions are classified as \textit{Insignificant}, most commonly because the apparent peak is significant in only a small fraction of the sample or forms part of broader detector-scale structure.

Fig.~\ref{fig:example_significant_residual_regions} illustrates the nine adopted classifications. Panels A--C show three forms of stellar-modelling mismatch: a line that is too strong in the synthesis, a line that is too weak, and a structured profile mismatch. Panels D--F show a low-recurrence or detector-scale feature, residual sky emission, and interstellar absorption. Panels G--I illustrate progressively less certain identifications: an unidentified feature, a known stellar line absent from the GALAH DR4 line list, and an ambiguous blend for which several explanations remain plausible.

Fe~{\sc i} accounts for the largest number of candidate line-strength mismatches, including 41 of the 130 \textit{$\log gf$ too high} regions and 36 of the 94 \textit{$\log gf$ too low} regions. Recurrent classifications (too high/low) are also found for Ti~{\sc i} (11/10), Ce~{\sc ii} (11/9), Cr~{\sc i} (9/5), Si~{\sc i} (13/0), V~{\sc i} (8/0), Ca~{\sc i} (8/0), and Ni~{\sc i} (6/0). These counts are descriptive rather than occurrence rates because the number and strength of available transitions vary substantially between species. The eight \textit{Known missing} features comprise four Cr~{\sc i}, three Fe~{\sc i}, and one Si~{\sc i} transition.

The classifications represent the most plausible explanation supported by the available diagnostics rather than unique physical identifications. In particular, the categories \textit{$\log gf$ too high} and \textit{$\log gf$ too low} indicate the direction in which the synthesised line strength would need to change. The underlying cause could instead involve an unresolved blend, inaccurate abundance, incomplete hyperfine or isotopic structure, wavelength or broadening errors, or deficiencies in the adopted line-formation treatment. The \textit{Line profile}, \textit{Unknown line}, and \textit{Unclear} classes explicitly identify cases for which a simple adjustment of the oscillator strength would be insufficient or poorly justified.

A complete version of Tab.~\ref{tab:tabular_peak_information}, including the classifications, supporting measurements, and inspection comments for all 380 regions, is provided as a machine-readable FITS table in the public repository\footnote{\url{https://github.com/svenbuder/residual_cannon/blob/main/data/galah_dr4_residual_peaks.fits}}. The corresponding diagnostic figure for every region is also provided in the repository.

\section{Data-driven extraction of residual information}
\label{sec:analysis_cannon}

The audit established that the residuals contain coherent stellar, interstellar, and observational structure. We now ask whether this mismatch is predictive by applying the data-driven machinery of \textsc{The Cannon} to residual rather than observed spectra. We test labels already included in the GALAH fit, unfitted extinction and activity indicators, and the relative information in wavelength pixels used or excluded during the GALAH DR4 optimisation.

\subsection{Residual modelling and validation with \textsc{The Cannon}}
\label{sec:analysis_cannon_method}

We model the residual spectra using \textsc{The Cannon} \citep{Ness2015} with an updated version\footnote{We use v.0.9.00 from \url{https://github.com/ANU-RSAA/AnniesLasso}, maintained by Marc White at the ANU Research School of Astronomy and Astrophysics.} of the code by \citet{Casey2016}. Unlike its conventional application, in which \textsc{The Cannon} learns the relationship between observed flux and stellar labels, here we train it on the residual flux between each observed spectrum and its best-fitting synthetic spectrum.

For a star $n$ at pixel $\lambda$, we model each residual $\Delta f_{n\lambda}$ as
\begin{equation}
\Delta f_{n\lambda}
=
\boldsymbol{v}(\boldsymbol{\ell}_n) \cdot \boldsymbol{\theta}_{\lambda} + \epsilon_{n\lambda},
\label{eq:cannon_residual_model}
\end{equation}
where $\boldsymbol{\ell}_n$ contains the selected labels of star $n$, $\boldsymbol{v}(\boldsymbol{\ell}_n)$ is the corresponding design vector, and $\epsilon_{n\lambda}$ is a noise term drawn from a Gaussian distribution whose variance combines the observational flux uncertainty $\sigma_{n\lambda}$ and an additional wavelength-dependent model scatter $\sigma_{\mathrm{model},\lambda}$,
\begin{equation}
\epsilon_{n\lambda} \sim \mathcal{N}\left(0, \sigma_{n\lambda}^{2}+\sigma_{\mathrm{model},\lambda}^{2}\right).
\end{equation}

Throughout this work, we use a quadratic model, for which the design vector contains a constant term, the linear and quadratic terms of each label, and all pairwise cross-terms. The wavelength-dependent coefficient vector $\boldsymbol{\theta}_{\lambda}$ therefore describes how the residual flux changes across label space. For numerical stability, \textsc{The Cannon} internally centres and scales the input labels before constructing the quadratic design vector.

We train the model independently at each wavelength pixel using stars with known reference labels to determine the coefficient vector $\boldsymbol{\theta}_{\lambda}$ and model scatter $\sigma_{\mathrm{model},\lambda}$. At each wavelength, the implementation minimises the inverse-variance-weighted $\chi^2$ in the training residuals using the L-BFGS-B algorithm implemented in \texttt{scipy.optimize.minimize}, with Powell optimisation as a fallback. The model scatter is subsequently estimated by requiring the median normalised squared residual to be unity. After this training step, we fix these quantities and infer the labels of any residual spectrum by minimising the summed $\chi^2$ across all wavelength pixels, equivalent to maximising the Gaussian likelihood. In practice, the code minimises the weighted residual sum of squares using \texttt{scipy.optimize.least\_squares} with the Levenberg-Marquardt algorithm \citep{Scipy}.

The models used in this section differ in their training sample, input labels, and wavelength coverage. They test whether residuals retain information about stellar parameters, abundance patterns, extinction, and activity, and whether that information is concentrated in pixels used or masked during the GALAH DR4 optimisation, as already listed in Tab.~\ref{tab:residual_models}.

We analysed the robustness of our data-driven extraction through five-fold cross-validation. Using the \textsc{KFold} implementation of \textsc{sklearn}'s \textsc{model\_selection} \citep{scikit-learn}, we divided each sample into five reproducibly shuffled subsets. For each split, we trained the model on $80\,\mathrm{\%}$ of the spectra and evaluated it on the remaining $20\,\mathrm{\%}$, such that every spectrum was used for testing once.

We quantify the recovery of each label using
\begin{equation}
\Delta \ell
=
\ell_{\mathrm{Cannon}}
-
\ell_{\mathrm{reference}}.
\label{eq:label_residual}
\end{equation}

\begin{figure}
    \centering
    \includegraphics[width=0.49\linewidth]{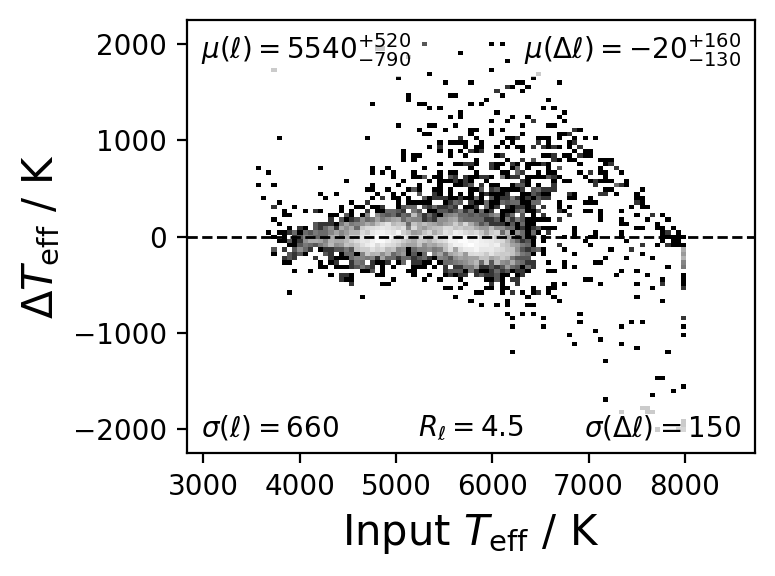}
    \includegraphics[width=0.49\linewidth]{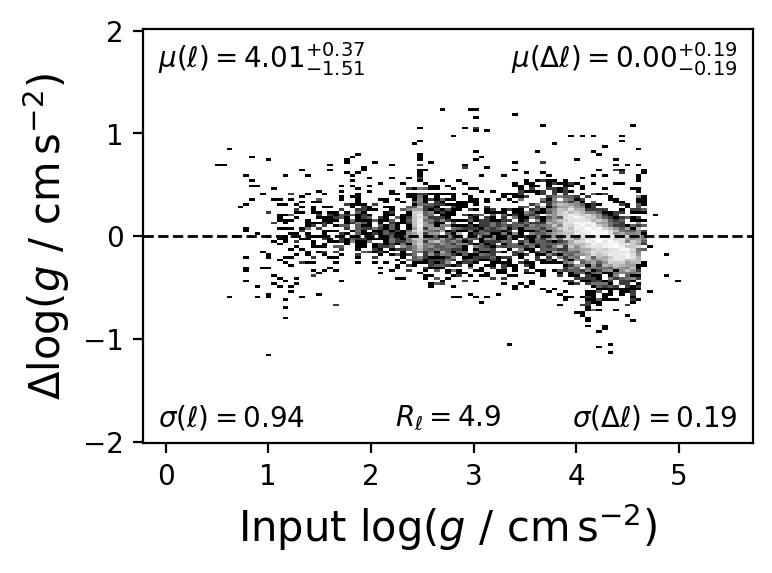}
    \includegraphics[width=0.49\linewidth]{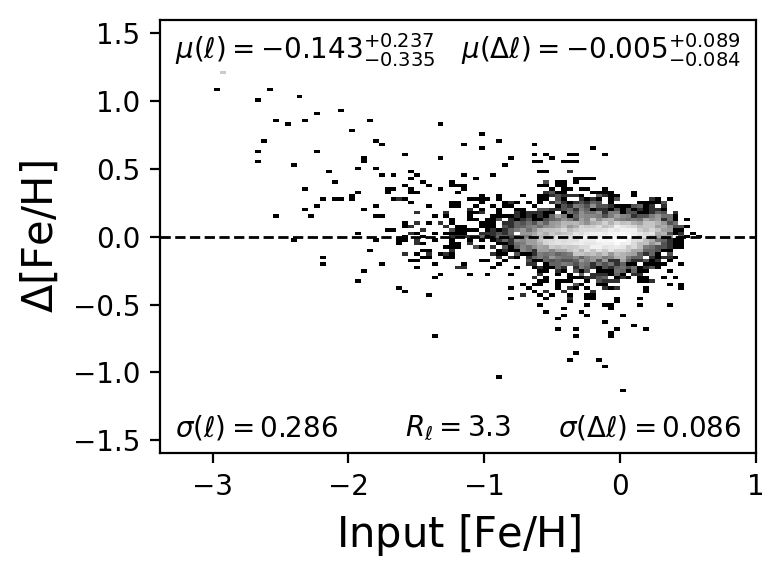}
    \includegraphics[width=0.49\linewidth]{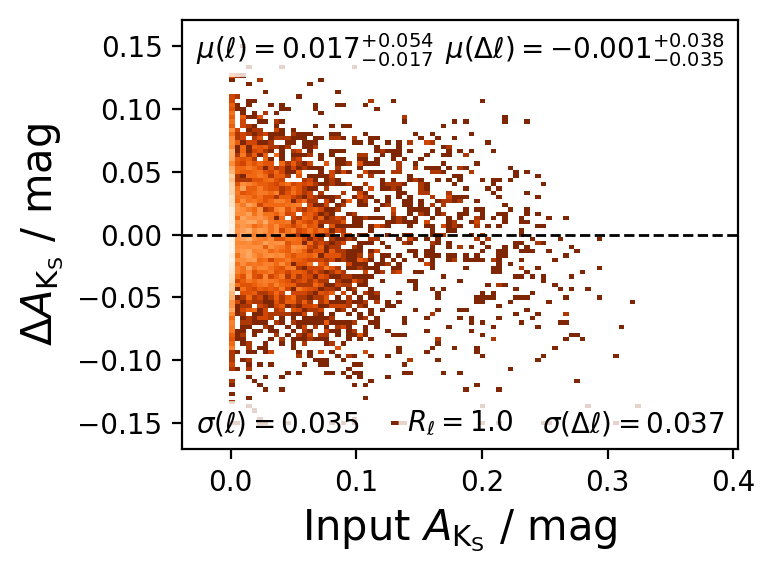}
    \includegraphics[width=0.49\linewidth]{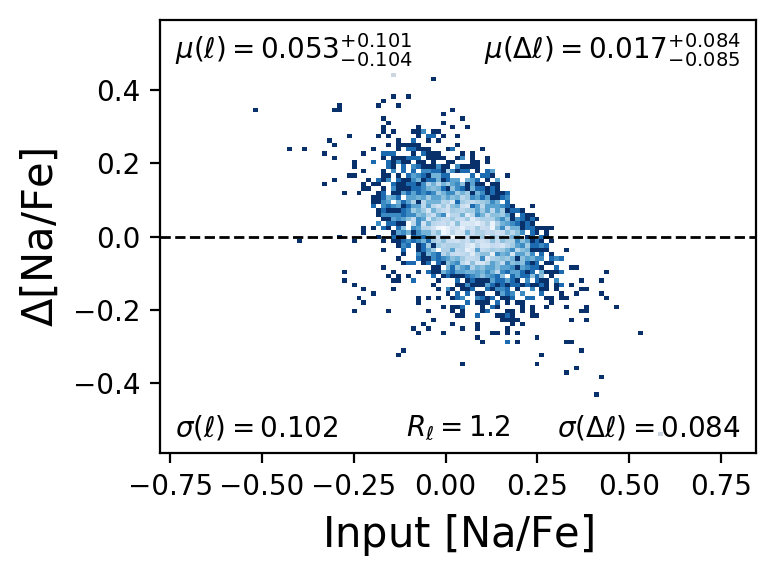}
    \includegraphics[width=0.49\linewidth]{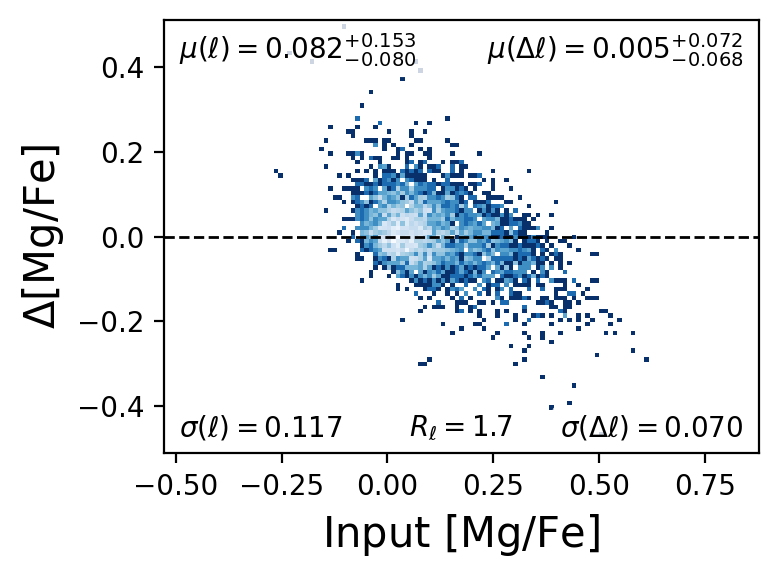}
    \includegraphics[width=0.49\linewidth]{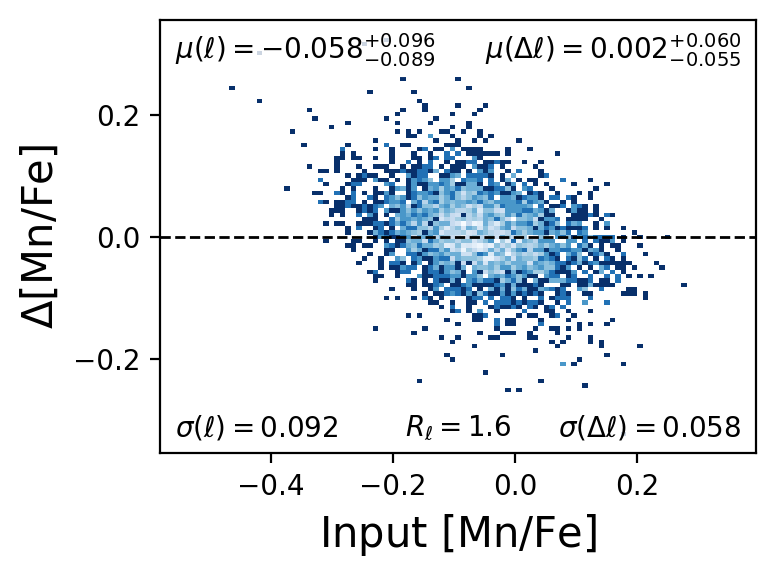}
    \includegraphics[width=0.49\linewidth]{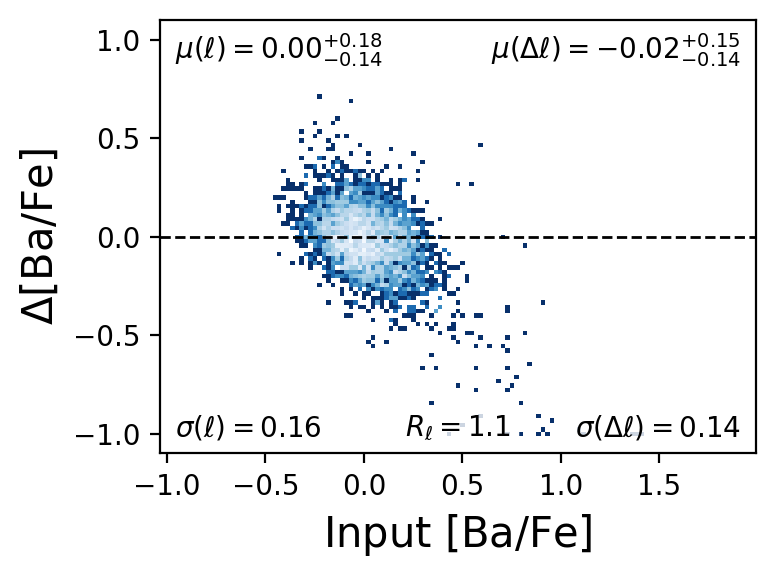}
    \caption{Cross-validated recovery differences, $\Delta\ell=\ell_{\mathrm{Cannon}}-\ell_{\mathrm{reference}}$, as a function of the reference labels for Models~\textcolor{M1}{M1} (grey), \textcolor{M2}{M2} (blue), and \textcolor{M3}{M3} (orange). From top left to bottom right, the panels show $T_\mathrm{eff}$, $\log g$, $[\mathrm{Fe/H}]$, $A_{K_\mathrm{S}}$, $[\mathrm{Na/Fe}]$, $[\mathrm{Mg/Fe}]$, $[\mathrm{Mn/Fe}]$, and $[\mathrm{Ba/Fe}]$. Dashed lines mark $\Delta\ell=0$. Annotations give the 16th, 50th, and 84th percentiles of the label and recovery distributions and the label-recovery ratio $R_\ell$. Equal axis scaling in label units allows the recovery scatter to be compared directly with the label spread.
    }
    \label{fig:recovery_comparison}
\end{figure}

\begin{figure*}
    \centering
    \includegraphics[width=\linewidth]{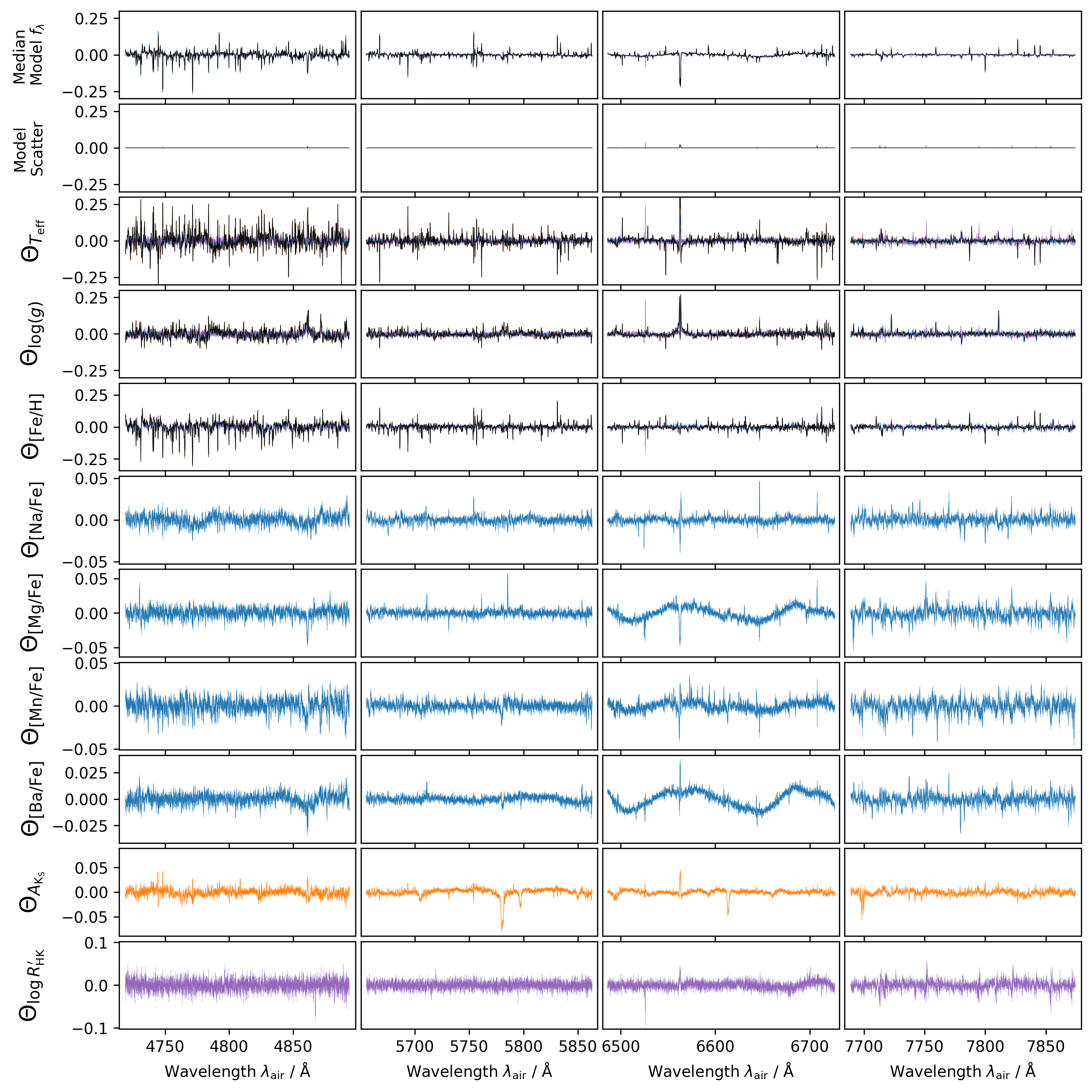}
    \caption{Selected coefficients across wavelength for Models~\textcolor{M1}{M1}, \textcolor{M2}{M2}, \textcolor{M3}{M3}, and \textcolor{M4}{M4}. Lines show the median coefficient across the five cross-validation folds, and shaded regions show the full fold-to-fold range. Because the labels are internally centred and scaled, the coefficients describe the local residual response near the centre of each training-label distribution.}
    \label{fig:cannon_linear_coefficients_joint}
\end{figure*}

Figures~\ref{fig:recovery_comparison}, \ref{fig:recovery_comparison2}, and \ref{fig:recovery_comparison3} show the recovery $\Delta \ell$ over $\ell$ for selected labels.\footnote{Plots for all remaining labels are available in our \href{https://github.com/svenbuder/residual_cannon}{repository}.} For each panel, we also calculated robust standard deviations from the 16th--84th percentile intervals both of the labels and their recovery. We further quantify the recovery spread relative to the sampled label spread by using

\begin{equation}
    R_\ell = \frac{\sigma(\ell)}{\sigma(\Delta \ell)}.
\end{equation}
Larger $R_\ell$ indicates that prediction errors are small relative to the label range.
These performance indicators, in particular a higher ratio $R_\ell$, establish whether the residuals contain predictive information. The wavelength-dependent linear coefficients, corresponding to the first derivatives of the model with respect to each scaled label (see Figs.~\ref{fig:cannon_linear_coefficients_joint} and \ref{fig:cannon_linear_coefficients_model5}), are then used to identify where that information resides and to compare the learned residual sensitivity with known spectral lines and the residual catalogue from Sec.~\ref{sec:analysis_peak_catalogue}. For each wavelength pixel, we show the median coefficient across the five folds together with its full fold-to-fold range. Because each fold contains 80\% of the sample, the internal label scaling and resulting coefficients differ only slightly. The narrow envelopes for almost all linear coefficients therefore demonstrate that the learned wavelength dependence is robust to the choice of training subset; they are not interpreted as formal coefficient uncertainties.

\subsection{Information about fitted stellar labels}
\label{sec:analysis_cannon_fitted_labels}

The GALAH synthesis was explicitly optimised using global stellar parameters and individual elemental abundances. As a reference scale, the median uncertainties reported by GALAH DR4 for stars in the \textit{abundance} sample are $70\,\mathrm{K}$ in $T_\mathrm{eff}$, $0.10\,\mathrm{dex}$ in $\log g$, and $0.060\,\mathrm{dex}$ in $[\mathrm{Fe/H}]$, which all include absolute uncertainties as well. For the elemental abundances, the corresponding median precision uncertainties are $0.035\,\mathrm{dex}$ for $[\mathrm{Na/Fe}]$, $0.029\,\mathrm{dex}$ for $[\mathrm{Mg/Fe}]$, $0.031\,\mathrm{dex}$ for $[\mathrm{Mn/Fe}]$, and $0.068\,\mathrm{dex}$ for $[\mathrm{Ba/Fe}]$.

Recovering these labels from the residuals would therefore show that the remaining model-data mismatch still varies systematically across the label space used in the original fit. We first examine the global stellar parameters in Sec.~\ref{sec:analysis_cannon_fitted_stellar_parameters}, before testing for more specific information about elemental abundance patterns in Sec.~\ref{sec:analysis_cannon_fitted_abundances}.

\subsubsection{Stellar parameters}
\label{sec:analysis_cannon_fitted_stellar_parameters}

Model \textcolor{M1}{M1} tests the recovery of $T_\mathrm{eff}$, $\log g$, and $[\mathrm{Fe/H}]$ using the baseline sample, which spans the broadest stellar-parameter range considered here. Model \textcolor{M1}{M1*} repeats the same experiment using the more restricted abundance sample, which excludes the hottest stars and applies additional quality criteria. This comparison tests whether the predictive residual structure persists within a more homogeneous, higher-quality sample, although it does not isolate the effects of signal-to-noise ratio, quality selection, and stellar-type coverage individually.

The cross-validated recovery results are combined in Fig.~\ref{fig:recovery_comparison} using model-specific colour maps; here, we focus on the grey-scale panels for Model~\textcolor{M1}{M1}. For this model, the robust recovery scales are $\sigma(\Delta T_\mathrm{eff})=150\,\mathrm{K}$, $\sigma(\Delta\log g)=0.19\,\mathrm{dex}$, and $\sigma(\Delta[\mathrm{Fe/H}])=0.086\,\mathrm{dex}$. These are substantially smaller than the corresponding label spreads of $660\,\mathrm{K}$, $0.94\,\mathrm{dex}$, and $0.286\,\mathrm{dex}$, yielding label-recovery ratios of $R_{T_\mathrm{eff}}=4.5$, $R_{\log g}=4.9$, and $R_{[\mathrm{Fe/H}]}=3.3$. The median biases are small: $-20\,\mathrm{K}$, $0.00\,\mathrm{dex}$, and $-0.005\,\mathrm{dex}$, respectively.

The recovery is not uniform across the sampled label space. For $T_\mathrm{eff}$, the scatter and frequency of outliers increase among the comparatively sparse stars hotter than approximately $6500\,\mathrm{K}$. The $\log g$ recovery shows a mildly asymmetric trend among dwarfs near $\log g\sim4$ and increased scatter around the red clump near $\log g\sim2.5$. The metallicity recovery is strongest across the well-sampled central range and deteriorates towards the metal-poor edge, where few training stars are available. The quoted global recovery scales should therefore be interpreted as cross-validated interpolation performance within the baseline sample, rather than as uniform precision across its full parameter range.

Model \textcolor{M1}{M1*} confirms that the residual information persists in the more restricted abundance sample. Its robust recovery scales are $85\,\mathrm{K}$ in $T_\mathrm{eff}$, $0.14\,\mathrm{dex}$ in $\log g$, and $0.059\,\mathrm{dex}$ in $[\mathrm{Fe/H}]$, compared with label spreads of $623\,\mathrm{K}$, $0.95\,\mathrm{dex}$, and $0.270\,\mathrm{dex}$. The corresponding ratios increase to $R_{T_\mathrm{eff}}=7.4$, $R_{\log g}=6.7$, and $R_{[\mathrm{Fe/H}]}=4.6$. The median biases remain negligible at $+0\,\mathrm{K}$, $0.00\,\mathrm{dex}$, and $-0.002\,\mathrm{dex}$. Model \textcolor{M1}{M1*} therefore performs better in both absolute terms and relative to the label coverage of its sample, although the combined sample-selection differences prevent us from attributing this improvement to any single factor. The complete recovery statistics are summarised in Tab.~\ref{tab:recovery_performance}.

\begin{table*}
    \centering
    \caption{Cross-validated recovery of $T_\mathrm{eff}$, $\log g$, and $[\mathrm{Fe/H}]$ for the residual models. Columns give the median recovery difference and its 16th--50th and 50th--84th percentile intervals, calculated from the combined left-out predictions of the five folds. \textcolor{M1}{M1*} denotes training of model \textcolor{M1}{M1} on the \textit{abundance} sample, rather than the \textit{baseline} sample.}   
    \label{tab:recovery_performance}
    \begin{tabular}{cccccccccc}
    \hline \hline
    Model 
    & \multicolumn{3}{c}{$\Delta T_\mathrm{eff}\,/\,\mathrm{K}$}
    & \multicolumn{3}{c}{$\Delta\log g\,/\,\mathrm{dex}$}
    & \multicolumn{3}{c}{$\Delta[\mathrm{Fe/H}]\,/\,\mathrm{dex}$} \\
    & $P_{50}$ & $P_{16}-P_{50}$ & $P_{a84}-P_{50}$
    & $P_{50}$ & $P_{16}-P_{50}$ & $P_{84}-P_{50}$
    & $P_{50}$ & $P_{16}-P_{50}$ & $P_{84}-P_{50}$ \\
    \hline
    \textcolor{M1}{M1}  & $-20$  & $-130$  & $+160$  & $+0.00~~$ & $-0.19~~$ & $+0.19~~$ & $-0.005$  & $-0.084$  & $+0.089$  \\
    \textcolor{M1}{M1*} & $+~~0$ & $-~~77$ & $+~~92$ & $+0.00~~$ & $-0.14~~$ & $+0.14~~$ & $-0.002$  & $-0.058$  & $+0.059$ \\
    \textcolor{M2}{M2}  & $+~~0$ & $-~~78$ & $+~~82$ & $-0.01~~$ & $-0.14~~$ & $+0.15~~$ & $-0.004$  & $-0.060$  & $+0.058$  \\
    \textcolor{M3}{M3}  & $-20$  & $-130$  & $+150$  & $+0.01~~$ & $-0.19~~$ & $+0.19~~$ & $-0.004$  & $-0.085$  & $+0.087$  \\
    \textcolor{M4}{M4}  & $+10$  & $-100$  & $+130$  & $-0.01~~$ & $-0.09~~$ & $+0.13~~$ & $+0.007$  & $-0.079$  & $+0.091$  \\
    \textcolor{M5}{M5}  & $-60$  & $-140$  & $+180$  & $-0.00~~$ & $-0.23~~$ & $+0.20~~$ & $-0.01~~$ & $-0.10~~$ & $+0.09~~$ \\
    \textcolor{M6}{M6}  & $+10$  & $-160$  & $+220$  & $+0.02~~$ & $-0.21~~$ & $+0.23~~$ & $+0.00~~$ & $-0.10~~$ & $+0.13~~$ \\    
    \hline \hline
    \end{tabular}
\end{table*}

The coefficient spectra panels in Fig.~\ref{fig:cannon_linear_coefficients_joint}, each showing the full range of coefficients across the five folds, identify the wavelength regions associated with this predictive information. Because the model is quadratic and the labels are internally centred and scaled, the plotted linear coefficients describe the local residual response near the centre of the training-set label distribution, conditional on the other labels. They should not be interpreted as global derivatives across the full parameter range; quadratic and cross-terms may become important elsewhere in label space.

The $T_\mathrm{eff}$ coefficient contains numerous narrow positive and negative features across all four wavelength arms, indicating that the temperature dependence is distributed across many residual spectral features. The strongest response occurs around $\mathrm{H}\upalpha$, where the line-core and wing residuals vary in opposite directions with temperature. This demonstrates a systematic temperature dependence in the remaining $\mathrm{H}\upalpha$ mismatch, but does not by itself distinguish between deficiencies in line formation, broadening, continuum normalisation, activity, or other modelling effects. A lower-amplitude, broader response is also visible across parts of the blue arm, potentially reflecting the combined influence of many weaker features.

The $\log g$ coefficient is likewise dominated by structure around $\mathrm{H}\upalpha$ and $\mathrm{H}\upbeta$, together with narrower responses associated with individual metal-line residuals. These include regions near the modelled Fe~{\sc i} lines at 4870.97 and 4891.16\,\AA\ with too high $\log gf$ and the candidate missing Fe~{\sc i} line near 7780.56\,\AA. Parts of the blue-arm response have the opposite sign to the temperature coefficient, illustrating the partial covariance of the residual dependences on $T_\mathrm{eff}$ and $\log g$.

The $[\mathrm{Fe/H}]$ coefficient contains many line-localised responses, consistent with metallicity information distributed across absorption features and blends that are not reproduced perfectly by the synthesis. Not every Fe-associated peak from Sec.~\ref{sec:analysis_peak_catalogue} appears prominently in the linear metallicity coefficient. Some features may instead vary non-linearly with metallicity, enter through cross-terms with $T_\mathrm{eff}$ or $\log g$, or become informative only within a restricted region of label space.

Three related diagnostics should be distinguished. The cross-validated recovery scales quantify predictive performance for out-of-sample spectra, the inferred wavelength-dependent model scatter describes the typically small residual variation not captured by the fitted label model, and the fold-to-fold coefficient ranges measure the stability of the learned wavelength response against changes in the training subset. The coefficient envelopes in Fig.~\ref{fig:cannon_linear_coefficients_joint} are narrow for almost all linear terms, showing that the identified wavelength structure is highly reproducible across the five folds.

\subsubsection{Element abundance patterns}
\label{sec:analysis_cannon_fitted_abundances}

Having established that the residual spectra retain substantial information about the global stellar parameters, we next test whether they preserve more specific chemical information. Model~\textcolor{M2}{M2} extends Model~\textcolor{M1}{M1*} by fitting $[\mathrm{Na/Fe}]$, $[\mathrm{Mg/Fe}]$, $[\mathrm{Mn/Fe}]$, and $[\mathrm{Ba/Fe}]$ alongside $T_\mathrm{eff}$, $\log g$, and $[\mathrm{Fe/H}]$. Both models use the same \textit{abundance} sample, allowing their performance to be compared without changing the underlying stellar sample. The recovery of the three global labels remains nearly unchanged, with robust scales of approximately $80\,\mathrm{K}$ in $T_\mathrm{eff}$, $0.15\,\mathrm{dex}$ in $\log g$, and $0.059\,\mathrm{dex}$ in $[\mathrm{Fe/H}]$ (Tab.~\ref{tab:recovery_performance}).

In contrast to the strong recovery of these global labels, the individual abundance ratios are recovered only modestly relative to their intrinsic sample spreads. The abundance recovery is shown by the blue-density panels in Fig.~\ref{fig:recovery_comparison}. The robust recovery scales $\sigma(\Delta[\mathrm{X/Fe}])$ are $0.084\,\mathrm{dex}$ for Na, $0.070\,\mathrm{dex}$ for Mg, $0.058\,\mathrm{dex}$ for Mn, and $0.14\,\mathrm{dex}$ for Ba. These values are larger than the corresponding median GALAH DR4 precision uncertainties by factors of approximately $1.9-2.4$.

The abundance recovery must also be considered relative to the limited label spread of the sample. The robust spreads $\sigma(\mathrm{[X/Fe]})$ are $0.102\,\mathrm{dex}$ for Na, $0.117\,\mathrm{dex}$ for Mg, $0.092\,\mathrm{dex}$ for Mn, and $0.16\,\mathrm{dex}$ for Ba. The corresponding label-recovery ratios are therefore $R_{[\mathrm{Na/Fe}]}=1.2$, $R_{[\mathrm{Mg/Fe}]}=1.7$, $R_{[\mathrm{Mn/Fe}]}=1.6$, and $R_{[\mathrm{Ba/Fe}]}=1.1$. The residual spectra thus contain modest predictive information about Mg and Mn, whereas Na and Ba are recovered only marginally better than a constant prediction near the centre of their training-label distributions. The recovery also deteriorates towards sparsely sampled abundance extremes, where individual training folds contain relatively few chemically enhanced stars.

The linear coefficients (blue in Fig.~\ref{fig:cannon_linear_coefficients_joint}) suggest that this abundance information is not carried primarily by clear residual signatures of the corresponding elements. In particular, the Mg and Ba coefficients retain broad, wave-like structure across the red arm, while several abundance coefficients respond strongly around the Balmer lines. The Mn and Ba coefficients also show structure near the diffuse interstellar bands around $5780\,\text{\AA}$, $5797\,\text{\AA}$, and $6613\,\text{\AA}$. By contrast, no single set of element-specific absorption features clearly dominates the coefficient spectrum for each abundance.

A comparison with the residual catalogue confirms that the abundance coefficients are rarely element-localised. The strong Si~{\sc i} residual near $7800\,\text{\AA}$ produces no prominent Na, Mg, Mn, or Ba response, while only Mg and Mn coefficients near $4796.36$, $4829.0$, and $4871.0\,\text{\AA}$ coincide with candidate Ti~{\sc i}, Ni~{\sc i}, and Fe~{\sc i} features of the same element group. Most identified stellar residuals respond more strongly to $[\mathrm{Fe/H}]$. Model~\textcolor{M2}{M2} thus seems to exploit correlations of abundances with stellar populations, atmospheric parameters, extinction, observing conditions, or reduction and normalisation artefacts rather than direct sensitivity to the nominal element. These correlations are reproducible across the folds, but may also enter through quadratic or cross-terms and should not be interpreted as independent abundance diagnostics.

\subsection{Information about properties not fitted by the synthesis}
\label{sec:analysis_cannon_unfitted_labels}

The abundance results show that predictive recovery does not have to arise from direct sensitivity to the corresponding atomic features. We therefore turn to properties not included in the GALAH synthesis, for which independent spectral diagnostics can help distinguish physically meaningful responses from indirect or contaminating correlations. We first examine interstellar extinction and absorption in Sec.~\ref{sec:analysis_cannon_extinction_labels}, followed by chromospheric activity in Sec.~\ref{sec:analysis_cannon_activity_labels}.

\subsubsection{Interstellar extinction and absorption}
\label{sec:analysis_cannon_extinction_labels}

Model~\textcolor{M3}{M3} extends Model~\textcolor{M1}{M1} by adding the line-of-sight extinction $A_{K_\mathrm{S}}$ to $T_\mathrm{eff}$, $\log g$, and $[\mathrm{Fe/H}]$, using the same \textit{baseline} sample. Although extinction does not directly alter a continuum-normalised stellar spectrum through its broad wavelength dependence, it correlates with narrow absorption produced by intervening interstellar material, including diffuse interstellar bands (DIBs) and atomic K~{\sc i}.

The baseline sample is strongly concentrated towards low-extinction sightlines, with $A_{K_\mathrm{S}} = 0.017_{-0.017}^{+0.054}\,\mathrm{mag}$ and the orange-density panel in Fig.~\ref{fig:recovery_comparison} therefore probes a highly asymmetric label distribution with a hard lower boundary near zero. 
The robust recovery scale, $\sigma(\Delta A_{K_\mathrm{S}})=0.037\,\mathrm{mag}$, is comparable to the label spread of $\sigma(A_{K_\mathrm{S}})=0.035\,\mathrm{mag}$, yielding $R_{A_{K_\mathrm{S}}}=1.0$. For such a narrow and strongly skewed distribution, however, this ratio is not by itself a reliable measure of predictive information: most stars have negligible extinction, for which both the reference and recovered values naturally cluster near zero. The absolute recovery differences generally remain below $0.1\,\mathrm{mag}$, including for the more extinguished sightlines, and stars with $A_{K_\mathrm{S}}\sim0.2\,\mathrm{mag}$ are preferentially assigned larger recovered values. The residual spectra therefore retain useful information about extinction, particularly towards the higher-extinction tail, even though the global recovery ratio is compressed by the label distribution.

The linear $A_{K_\mathrm{S}}$ coefficient in Fig.~\ref{fig:cannon_linear_coefficients_joint} provides clearer evidence for the physical origin of this information. Its strongest negative responses coincide with the established DIBs near $5780.59$, $6613.66$, and $5797.19\,\text{\AA}$, with peak linear-coefficient amplitudes $\delta (\Delta f) / \delta A_\mathrm{K_S}$ of approximately $-0.078$, $-0.045$, and $-0.030$, respectively. A similarly strong response near $7698.96\,\text{\AA}$, with an amplitude of approximately $-0.046$, coincides with interstellar K~{\sc i}. Additional responses align with DIBs identified by \citet{Vogrincic2023} near $4762.57$, $5705.21$, $6493.11$, $6496.67$, $6530.17$, and $6592.53\,\text{\AA}$. The agreement in wavelength and profile morphology shows that Model~\textcolor{M3}{M3} extracts identifiable interstellar absorption rather than only broad correlations between extinction and the stellar sample.

A complete one-to-one correspondence with the DIB catalogue of \citet{Vogrincic2023} is not expected. Their analysis shifted spectra into a common interstellar rest frame and combined high-quality spectra in reddening bins, whereas our coefficients are inferred from individual spectra in the stellar rest frame and from a sample dominated by nearly unreddened sightlines. Differences between the stellar and interstellar radial velocities can therefore broaden or dilute weak ISM responses in our analysis. This likely contributes to the absence of clear coefficients at several previously identified DIBs, including the features near $5747.62$ and $5784.78\,\text{\AA}$, as well as numerous weaker catalogue entries.

Conversely, Model~\textcolor{M3}{M3} shows additional responses near $4726.40$, $4771.47$, $4823.13$, $5849.91$, $6535.89$, $6660.74$, and $6699.33\,\text{\AA}$. These correlations should not automatically be classified as DIBs, because an extinction label can also trace stellar-population, observing, or reduction-related structure. This is particularly apparent for the narrow or step-like responses near $7737.34$, $7741.97$, $7747.26$, and $7826.79\,\text{\AA}$. Their morphology differs from the smoother and broader response (above 9 pixels), for example around the $5780\,\text{\AA}$ DIB (covering 29 pixels), making an interstellar origin less likely despite their correlation with $A_{K_\mathrm{S}}$.

The K~{\sc i} response near $7699\,\text{\AA}$ is also more complex than the isolated DIBs. The stellar synthesis includes the stellar photospheric K~{\sc i} line, while the observed spectrum may contain an additional interstellar component at a different radial velocity. When the stellar and interstellar components overlap, part of the additional absorption may be absorbed into the fitted stellar potassium abundance; when they are offset, the interstellar component remains as asymmetric or split residual structure around the synthesised stellar line. The coefficient should therefore be interpreted as evidence for interstellar K~{\sc i}, but not as a simple isolated line profile. We note that this behaviour might apply to several interstellar absorption features that overlap with fitted stellar absorption features.

\subsubsection{Chromospheric activity}
\label{sec:analysis_cannon_activity_labels}

Model~\textcolor{M4}{M4} tests whether residual spectra retain information about chromospheric activity by fitting the activity indicator $\log R'_\mathrm{HK}$ alongside $T_\mathrm{eff}$, $\log g$, and $[\mathrm{Fe/H}]$ for the 226 stars in the \textit{activity} sample. Extreme positive flux residuals, several of which arise from imperfectly subtracted sky emission, can strongly influence a model trained on such a small sample. We therefore cap continuum-normalised fluxes above 1.2 at this value and assign them an uncertainty of the original flux value to strongly reduce their weight. As shown below, this treatment limits but does not eliminate correlations with sky-subtraction residuals.

\begin{figure}
    \centering
    \includegraphics[width=0.49\linewidth]{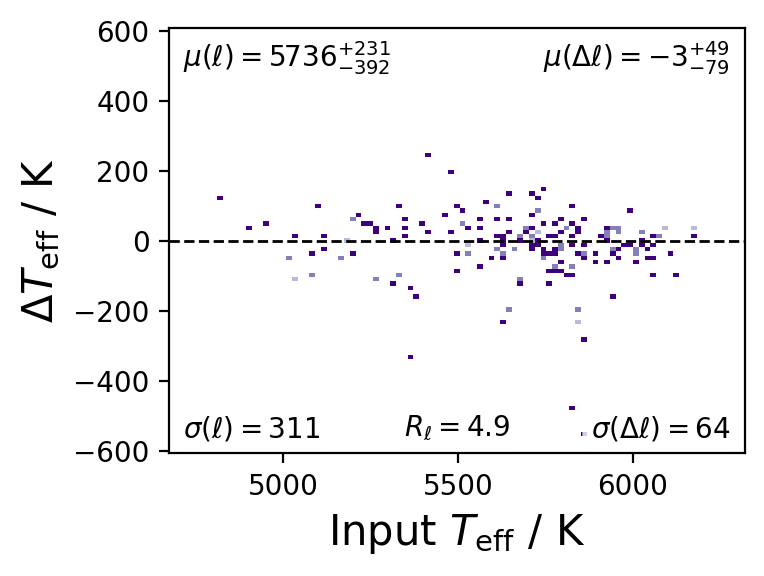}
    \includegraphics[width=0.49\linewidth]{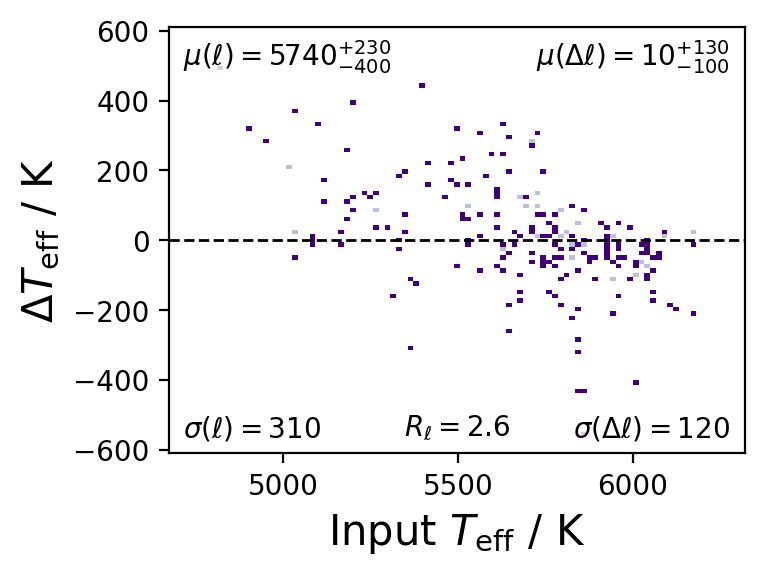}
    \includegraphics[width=0.49\linewidth]{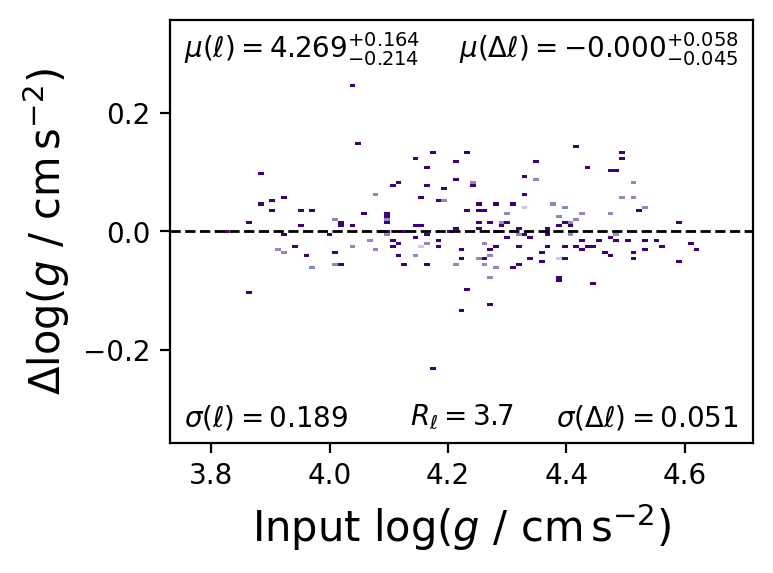}
    \includegraphics[width=0.49\linewidth]{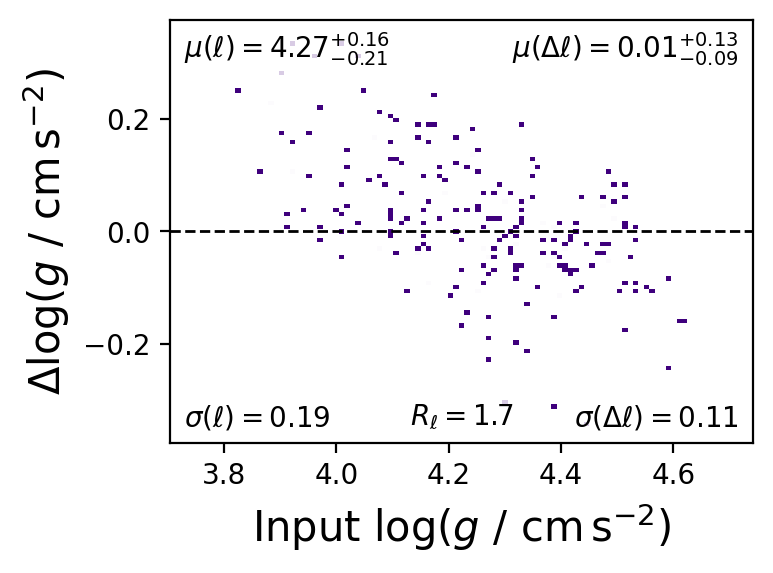}
    \includegraphics[width=0.49\linewidth]{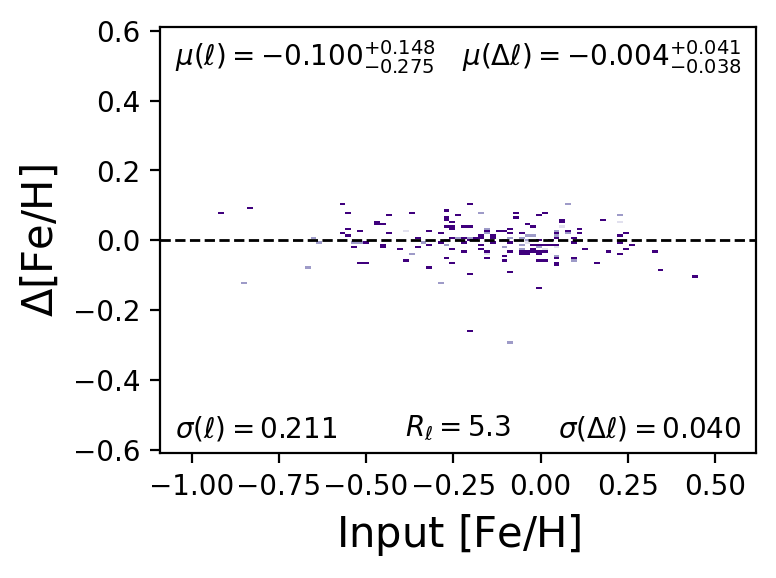}
    \includegraphics[width=0.49\linewidth]{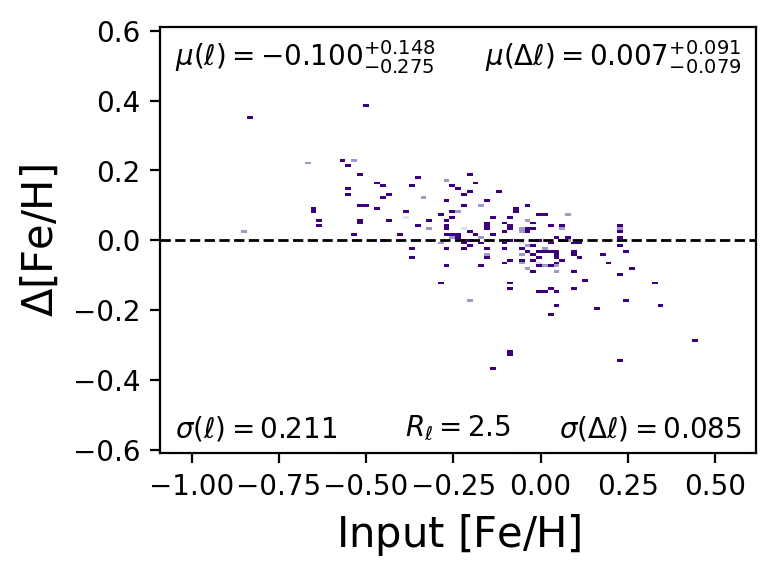}
    \includegraphics[width=0.49\linewidth]{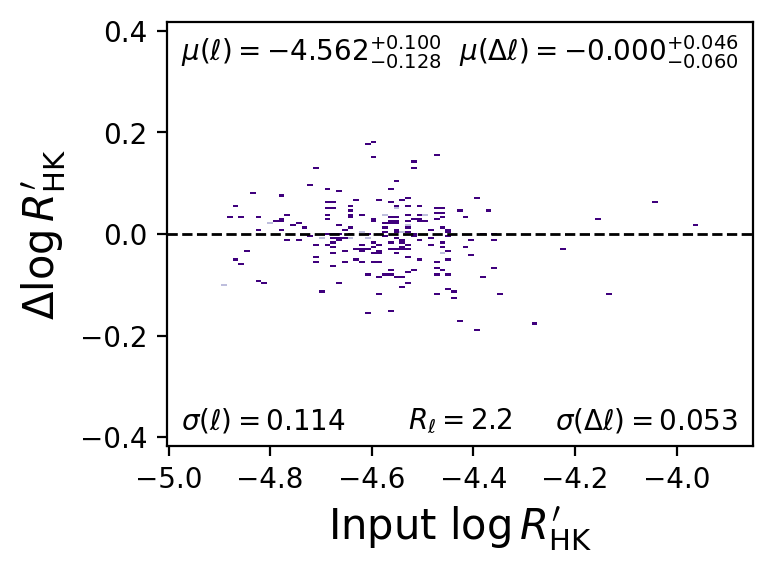}
    \includegraphics[width=0.49\linewidth]{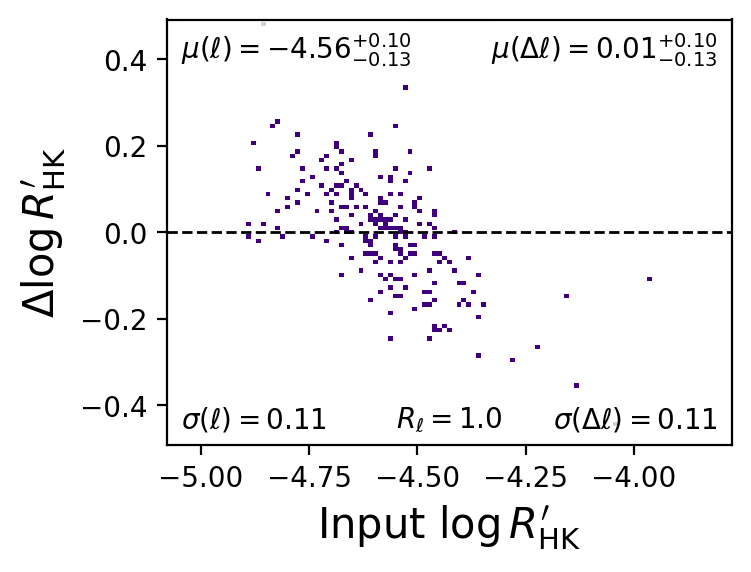}
    \caption{In-sample fit (left) and out-of-sample cross-validated recovery (right) of $T_\mathrm{eff}$, $\log g$, $[\mathrm{Fe/H}]$, and $\log R'_\mathrm{HK}$ with Model~\textcolor{M4}{M4} for the 226 stars in the \textit{activity} sample. The left panels show predictions from a model trained and evaluated on the complete sample and therefore do not provide an independent performance estimate. The right panels combine predictions for the left-out spectra from the five cross-validation folds. Plotting conventions follow Fig.~\ref{fig:recovery_comparison}.}
    \label{fig:recovery_comparison2}
\end{figure}

Fig.~\ref{fig:recovery_comparison2} contrasts the in-sample fit obtained when Model~\textcolor{M4}{M4} is trained and evaluated on all 226 spectra with its out-of-sample performance from five-fold cross-validation. The in-sample fit yields a robust recovery scale of $\sigma(\Delta\log R'_\mathrm{HK})=0.053$, compared with a label spread of $\sigma(\log R'_\mathrm{HK})=0.114$, corresponding to $R_{\log R'_\mathrm{HK}}=2.2$. This apparent recovery does not persist out of sample: the cross-validated recovery scale increases to $0.11$, giving $R_{\log R'_\mathrm{HK}}=1.0$. The diagonal trend in $\Delta\log R'_\mathrm{HK}$ indicates that the out-of-sample predictions regress towards the centre of the activity distribution rather than recovering its full range.

\begin{figure*}
    \centering
    \includegraphics[width=\linewidth]{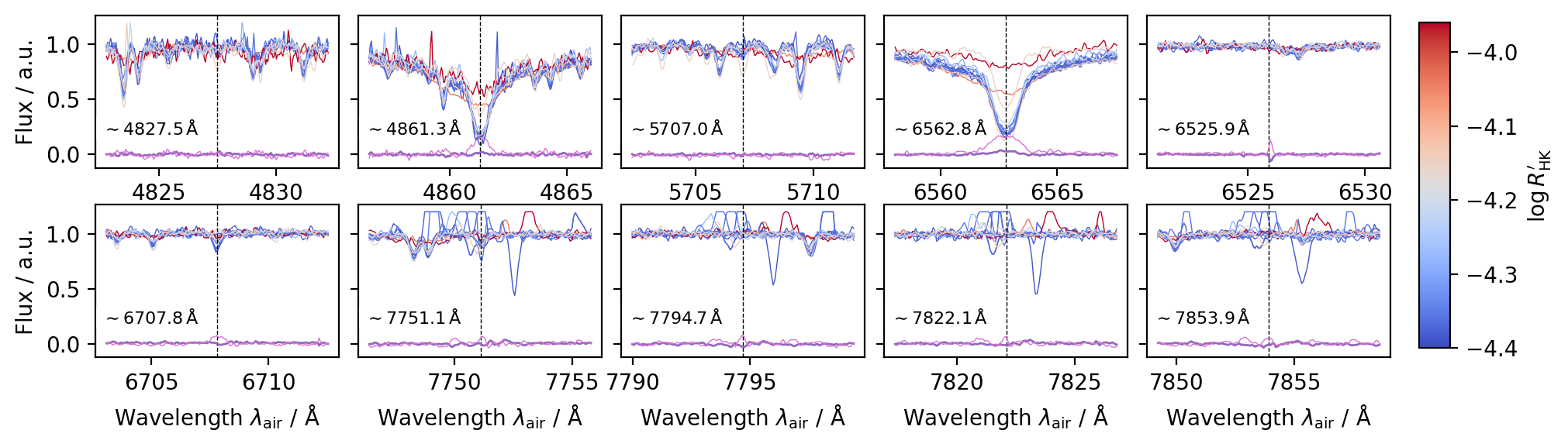}
    \caption{Observed flux for the 15 stars with highest $\log R'_\mathrm{HK}$ of the \textit{activity} sample as well as linear and quadratic coefficients of $\log R'_\mathrm{HK}$ (thick and thin purple lines) for selected wavelength regions.}
    \label{fig:logRHK_windows}
\end{figure*}

The precise in-sample fit does not generalise: with only approximately 180 training spectra per fold, the cross-validated predictions regress towards the centre of the activity distribution. This may reflect the small sample, uneven activity coverage, sensitivity to individual spectra, or correlations with unrelated residual features.

The coefficient spectra nevertheless provide evidence that the residuals contain some physically plausible activity-related structure. Fig.~\ref{fig:logRHK_windows} examines selected regions in greater detail by showing the observed spectra of the 15 stars with the largest $\log R'_\mathrm{HK}$ values together with the linear and quadratic activity coefficients at the bottom in dark and bright purple, respectively. The three most active stars (in redder colours) show particularly clear changes around $\mathrm{H}\upalpha$ and $\mathrm{H}\upbeta$, supporting an activity-related origin for the Balmer-line coefficients. They also differ around the Li~{\sc i} line at $6707.8\,\text{\AA}$, a region already identified as imperfectly represented by the GALAH DR4 neural-network synthesis \citep[Sec.~8.3.4]{Buder2025b}. Because lithium is primarily associated with stellar youth \citep[see e.g.][]{Zerjal2019}, rather than being a direct tracer of chromospheric emission, any remaining response is more plausibly interpreted as a combination of this known modelling limitation and the correlation between youth and activity within the sample.

Of the activity-sensitive lines identified by \citet{Wise2018} from HARPS spectra, four lie within the GALAH wavelength range: $4827.46$, $4861.33$, $5707.00$, and $6562.81\,\text{\AA}$ (see Fig.~\ref{fig:logRHK_windows}). The latter two Balmer lines, $\mathrm{H}\upbeta$ and $\mathrm{H}\upalpha$, show clear responses in the $\log R'_\mathrm{HK}$ coefficients, whereas the features near $4827.46$ and $5707.00\,\text{\AA}$ show no prominent response. Among the 32 HARPS lines studied by \citet{Lisogorskyi2019} within our wavelength range (attached in Fig.~\ref{fig:logRHK_lines_Lisogorskyi2019}), we also identify a signal near $4736.78$, $4859.75$, and $4890.76\,\text{\AA}$, in addition to the Balmer lines.

Not all strong coefficient responses have a stellar origin. Model~\textcolor{M4}{M4} also responds to residual sky-emission features, including those near $7751.1$, $7794.7$, $7822.1$, and $7853.9\,\text{\AA}$ (in the bottom right of Fig.~\ref{fig:logRHK_windows}). Some of these responses vary substantially between the cross-validation folds, indicating that they are driven by a small number of influential spectra rather than by a stable relation with chromospheric activity. Their presence illustrates that a reproducible correlation between a residual feature and an activity label does not by itself establish a causal connection: activity can correlate accidentally with observing conditions, reduction artefacts, or sample selection.

\begin{figure}
    \centering
    \includegraphics[width=0.49\linewidth]{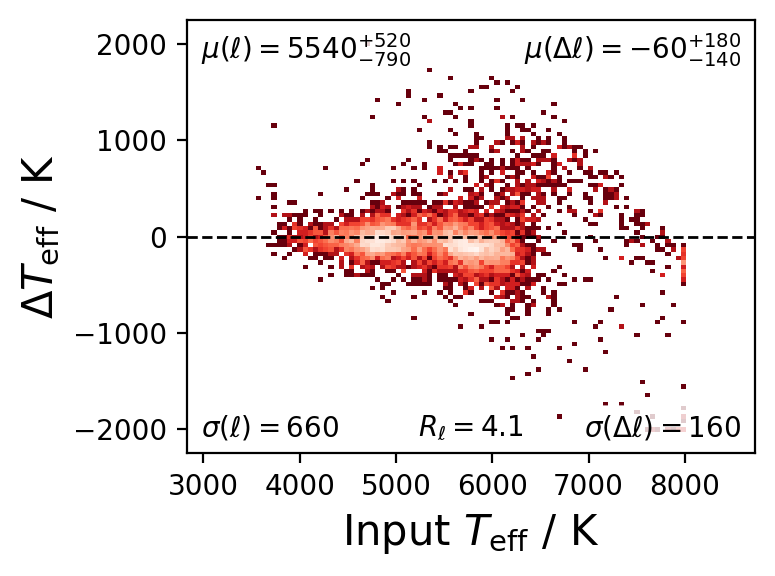}
    \includegraphics[width=0.49\linewidth]{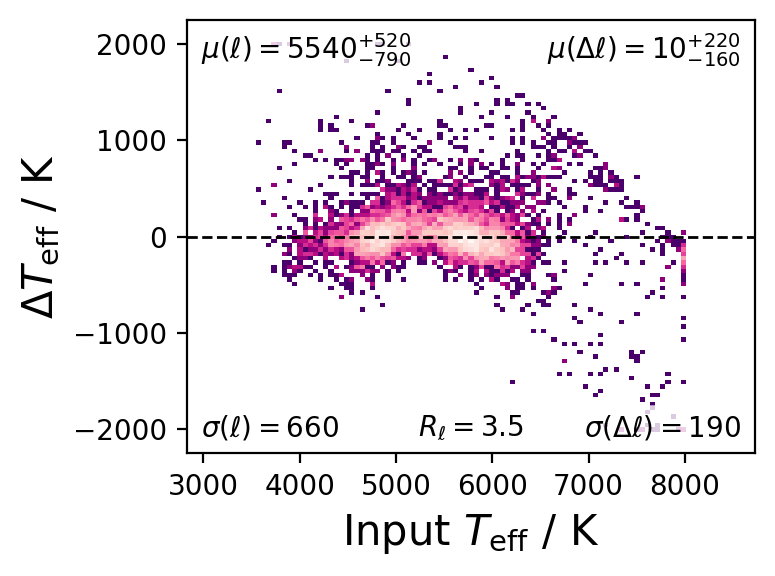}
    \includegraphics[width=0.49\linewidth]{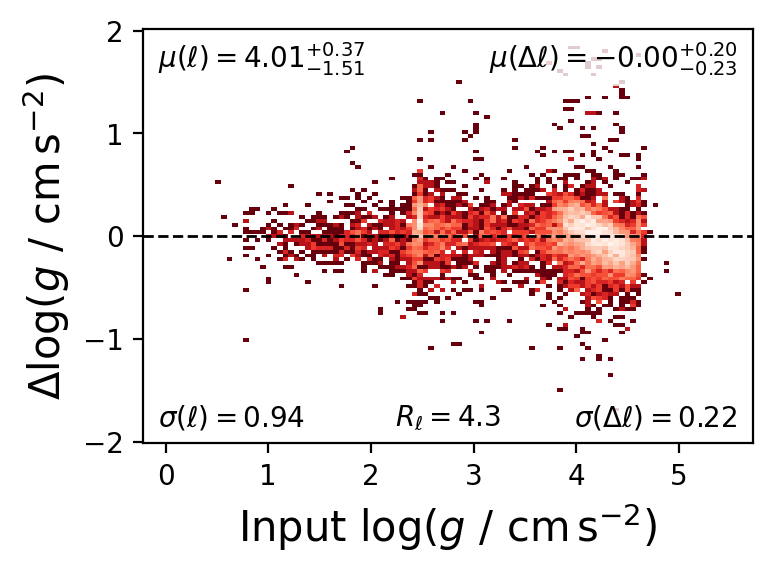}
    \includegraphics[width=0.49\linewidth]{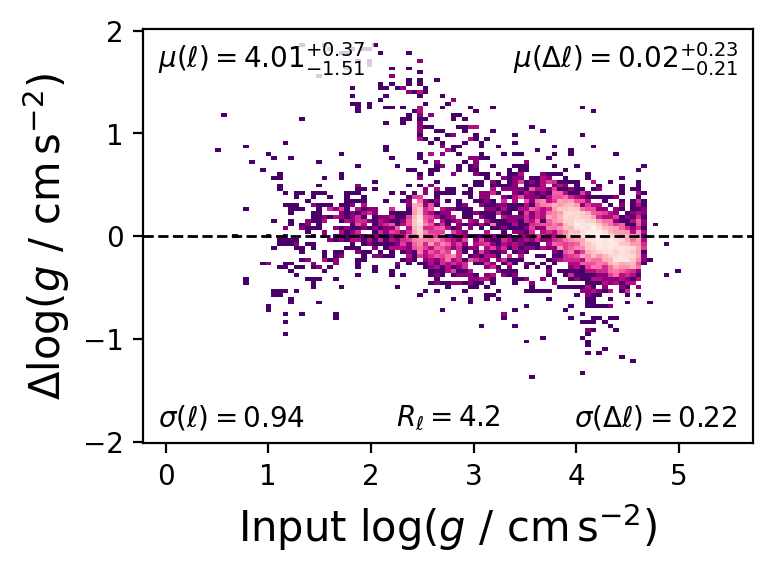}
    \includegraphics[width=0.49\linewidth]{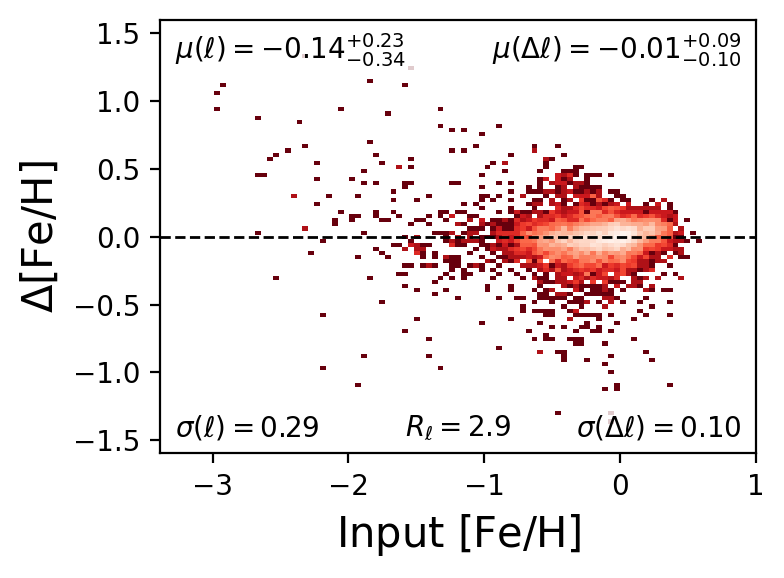}
    \includegraphics[width=0.49\linewidth]{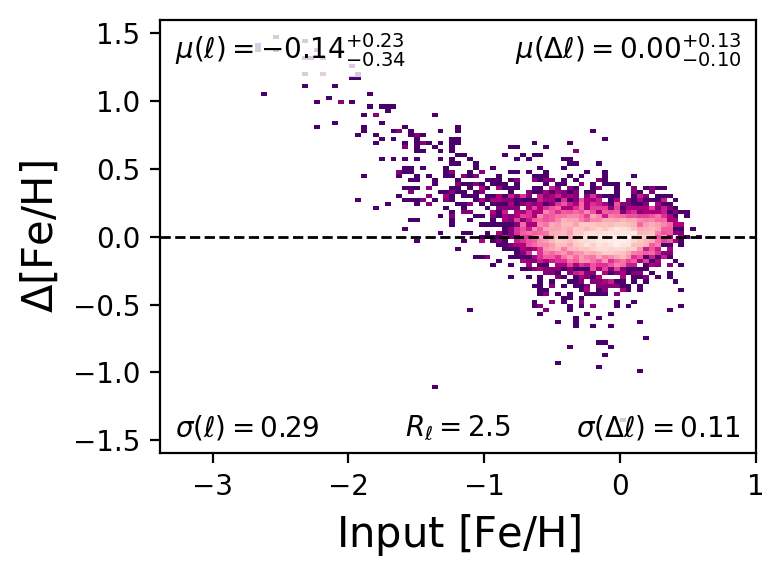}
    \caption{Cross-validated recovery of $T_\mathrm{eff}$, $\log g$, and $[\mathrm{Fe/H}]$ using the 793 pixels excluded from the GALAH DR4 optimisation with Model~\textcolor{M5}{M5} (left; red) and the 13134 pixels included in the optimisation with Model~\textcolor{M6}{M6} (right; pink). Plotting conventions follow Fig.~\ref{fig:recovery_comparison}.}
    \label{fig:recovery_comparison3}
\end{figure}

The coefficients nevertheless recover plausible Balmer and metal-line responses while also tracing unstable sky-emission residuals. The current model therefore demonstrates that activity-related information is present, but not that $\log R'_\mathrm{HK}$ can be recovered reliably. A larger, more diverse sample and more comprehensive masking of contaminated regions are required for robust out-of-sample prediction.

\begin{figure*}
    \centering
    \includegraphics[width=\linewidth]{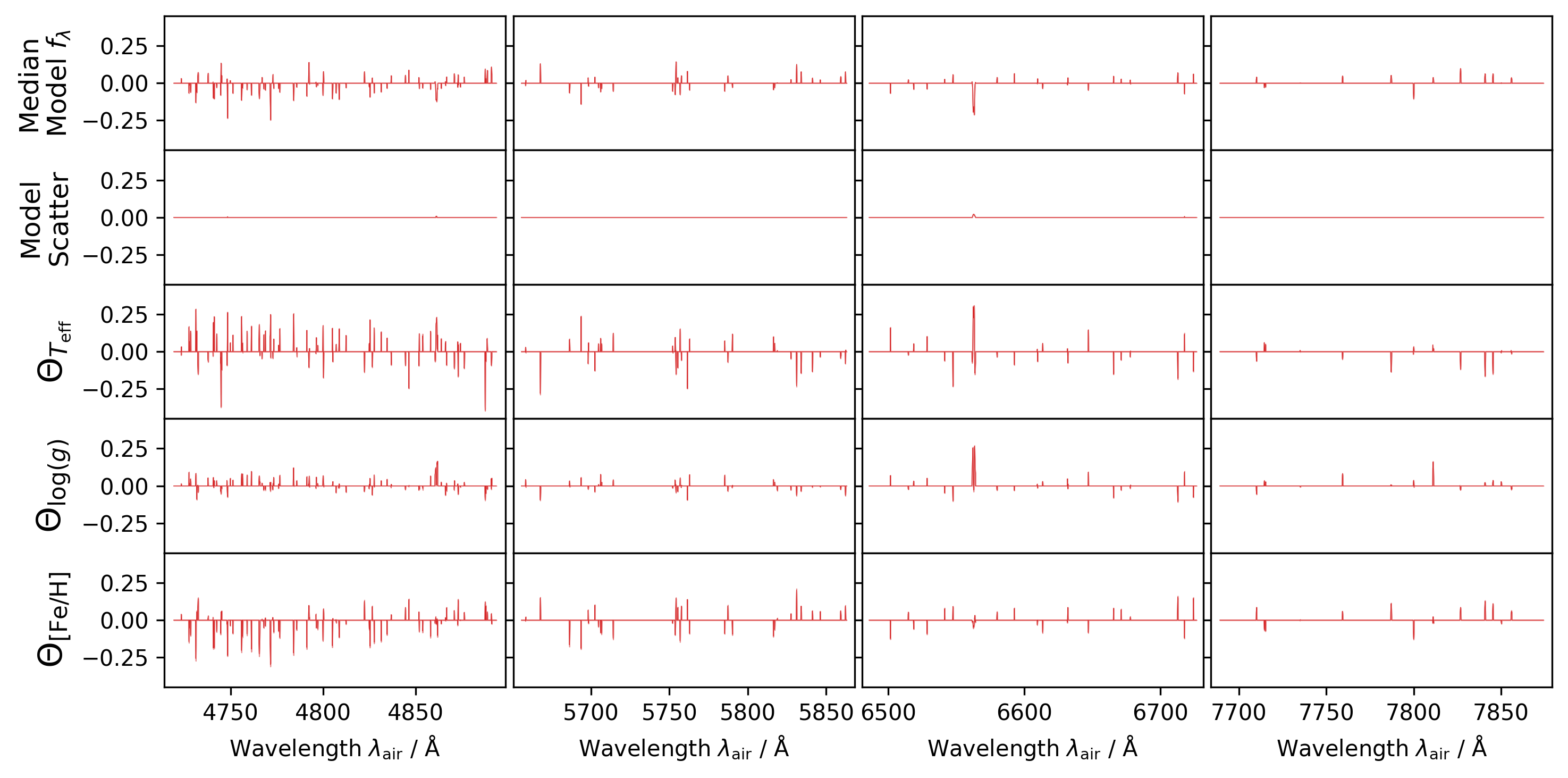}
    \caption{Selected coefficients across wavelength for Model~\textcolor{M5}{M5}, using only the 793 wavelength pixels excluded from the GALAH DR4 optimisation. Lines show the median coefficient across the five cross-validation folds, and shaded regions show the full fold-to-fold range.}
    \label{fig:cannon_linear_coefficients_model5}
\end{figure*}

\subsection{Where is the residual information concentrated?}
\label{sec:analysis_cannon_masked_pixels}

Of the 380 significant residual regions identified in Sec.~\ref{sec:analysis_peak_catalogue}, 111 ($29\,\mathrm{\%}$) fall within pixels excluded from the GALAH DR4 optimisation, while 269 ($71\,\mathrm{\%}$) fall within the retained wavelength range. This is a strong over-representation: the excluded region contains only 793 of the 13927 pixels ($6\,\mathrm{\%}$). Such a concentration is partly expected because these pixels were excluded during the DR4 analysis where representative spectra showed imperfect synthesis or reduction behaviour.

We now test whether this concentration of residual structure also corresponds to a concentration of recoverable stellar-label information. We divide the 13927 pixels into the 793 excluded from the GALAH DR4 optimisation and the 13134 retained pixels. Models~\textcolor{M5}{M5} and \textcolor{M6}{M6} fit the same labels, spectra, and cross-validation folds as Model~\textcolor{M1}{M1}, but use only the excluded and retained subsets, respectively.

Despite using only $6\,\mathrm{\%}$ of the wavelength pixels, Model~\textcolor{M5}{M5} recovers $T_\mathrm{eff}$, $\log g$, and $[\mathrm{Fe/H}]$ to approximately $160\,\mathrm{K}$, $0.22\,\mathrm{dex}$, and $0.095\,\mathrm{dex}$, only moderately worse than the full-spectrum values of $150\,\mathrm{K}$, $0.19\,\mathrm{dex}$, and $0.086\,\mathrm{dex}$ (see Fig.~\ref{fig:recovery_comparison3}). Model~\textcolor{M6}{M6}, using the remaining $94\,\mathrm{\%}$, reaches $190\,\mathrm{K}$, $0.22\,\mathrm{dex}$, and $0.12\,\mathrm{dex}$. The excluded pixels therefore outperform the 17-times-larger retained subset for all three labels.

This is not a random-subset comparison: the pixels were excluded precisely because their synthesis or reduction was less reliable and thus more likely to leave systematic residuals. Fig.~\ref{fig:cannon_linear_coefficients_model5} confirms that many narrow Model~\textcolor{M1}{M1} features remain in the excluded subset, whereas much of the broad Balmer structure lies outside its mask. Their narrow fold-to-fold ranges show that this information is reproducible across the cross-validation splits.

\section{Discussion}
\label{sec:discussion}

The two complementary analyses presented above lead to a consistent conclusion: the mismatch between observed and best-fitting synthetic spectra is neither random nor devoid of information. The systematic audit identifies recurrent residual structure with plausible stellar, interstellar, and observational origins, while the data-driven experiments show that this structure remains predictive of both fitted and unfitted properties. At the same time, the abundance and activity experiments demonstrate that predictive correlations are not necessarily direct or causal. We therefore discuss first what the residuals reveal about the limitations of current spectral modelling, then what can and cannot be inferred from their predictive information, and finally how these results can guide progress towards modelling a larger fraction of the observed spectrum.

\subsection{Residuals are systematic rather than noise}
\label{sec:discussion_systematic}

The excess, wavelength localisation, recurrence, and preferred sign of the residuals show that they cannot be described solely by independent noise or a uniform rescaling of the flux uncertainties. Instead, the residual field superposes narrow stellar-line mismatches, broad parameter-dependent Balmer structure, interstellar absorption, sky emission, and detector- or reduction-related patterns. Their signs indicate the direction in which the modelled flux must change, but not uniquely whether the cause is line opacity, blending, line formation, emission, normalisation, or reduction. The residuals must therefore be interpreted component by component rather than absorbed into a single additional noise term. Their overall prevalence also varies across stellar-label space: cool and metal-rich stars show the largest fractions of significantly mismatched pixels, consistent with their denser and stronger absorption spectra. In the coolest stars, visual inspection identifies molecular features, including CN, $\mathrm{C}_2$, and TiO, as important contributors. Accurate full-spectrum modelling is therefore especially challenging where severe line blending combines with remaining uncertainties in molecular line data, molecular equilibrium, atmospheric structure, and continuum placement.

\subsection{What the residual catalogue reveals about current spectral models}
\label{sec:discussion_residual_causes}

\begin{figure*}
\centering
\includegraphics[width=\linewidth]{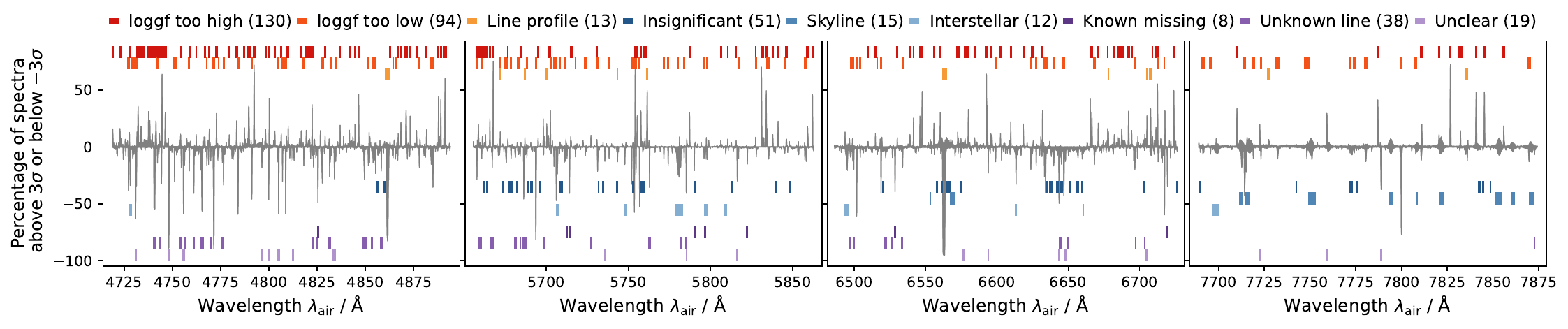}
\caption{Classification map of the recurrent signed residual structure across the four GALAH wavelength bands. The underlying curves reproduce the fractions of \textit{baseline}-sample spectra for which the observed flux is more than $3\sigma$ shallower or deeper than the synthesis, as shown in the bottom panels of Fig.~\ref{fig:residual_statistics}. Short coloured bars mark the 380 catalogue regions and are grouped vertically by their adopted primary classification: candidate line-strength and \textit{Line profile} mismatches in red and orange, \textit{Insignificant}, \textit{Skyline}, and \textit{Interstellar} regions in blue, and missing, unidentified, or ambiguous features in purple. The vertical offsets are categorical and carry no quantitative meaning; the horizontal bar widths are proportional to the fitted residual-region widths.}
\label{fig:residual_classification}
\end{figure*}

The classification of the 380 residual regions provides a direct overview of the origins of the recurrent signed structure. Fig.~\ref{fig:residual_classification} places the adopted classifications on top of the fractions of spectra with significantly deeper or shallower observed flux. This presentation links the residual catalogue back to the original statistical audit and demonstrates that similar residual amplitudes and signs can arise from physically different causes. The category assignments therefore require the additional atlas, line-list, interstellar, and sky-emission diagnostics described in Sec.~\ref{sec:analysis_peak_catalogue}.

The catalogue is dominated by candidate stellar line-strength and profile mismatches, while smaller subsets trace missing or unidentified lines, interstellar absorption, sky emission, and broader detector structure. This establishes stellar-line modelling as the largest identified contribution, although the classification sign indicates only the required change in total opacity and not necessarily an error in the oscillator strength itself.

These results highlight the continuing need for accurate experimental atomic data. Laboratory measurements provide the most reliable atomic parameters for stellar spectroscopy, yet many transitions of astrophysical importance still lack experimental determinations. Although individual residuals cannot be uniquely attributed to shortcomings in the atomic data alone, the prevalence of line-strength mismatches, profile discrepancies, and missing or unidentified features demonstrates that limitations in the underlying atomic data remain an important contributor to modelling errors. The residuals identified here demonstrate that improving the completeness and accuracy of atomic data remains a critical requirement for the next generation of stellar spectroscopic analyses.

\subsubsection{Line strengths, blends, and profiles}

Fe~{\sc i} contributes the largest absolute number of candidate line-strength mismatches, with 41 features classified as \textit{$\log gf$ too high} and 36 as \textit{$\log gf$ too low}. Substantial numbers are also associated with Ti~{\sc i}, Cr~{\sc i}, Si~{\sc i}, V~{\sc i}, Ca~{\sc i}, and Ce~{\sc ii}. These raw counts are descriptive rather than species-specific failure rates, because the number and strength of available transitions differ greatly between elements. Nevertheless, repeated mismatches identify groups of transitions for which the adopted line data warrant further investigation.

The most directly actionable cases are relatively isolated lines with a consistent residual sign. If a synthesised line is repeatedly too deep, its adopted opacity should be reduced; if it is too shallow, additional opacity is required. This does not uniquely identify the cause: an incorrect oscillator strength or blend contribution, abundance offset, or inadequately modelled line formation, including LTE/non-LTE effects, can produce the same residual direction. But it provides a well-defined direction for targeted line-list tests.

A particularly important example is the Si~{\sc i} feature near $7800\,\text{\AA}$, shown as the \textit{$\log gf$ too low} example in panel B of Fig.~\ref{fig:example_significant_residual_regions}. The currently adopted oscillator strength, $\log gf=-2.215$ \citep{K07}, produces substantially too little Si~{\sc i} absorption. Within the present synthesis, a value closer to $\log gf\approx-0.75$ \citep{Abia2020} reproduces the observed feature more closely. The mismatch is especially consequential because the Si~{\sc i} feature blends with the only Rb~{\sc i} line covered by GALAH. Underestimating the Si contribution can therefore bias or destabilise the decomposition of the blend and is the likely explanation for the currently imprecise measurements of Rb abundances in GALAH DR4. A similar issue was reported by \citet{Azhari2025}, who found that inaccurate Si~{\sc i} blending lines compromised the fit of a Th~{\sc ii} feature -- a rare radioactive tracer of rapid neutron-capture nucleosynthesis that can help constrain the role of events such as neutron-star mergers in producing the heaviest elements, while also enabling cosmic age dating \citep{Hill2002}.

Two related examples are the Na~{\sc i} line at $5688.19\,\text{\AA}$ and the Mg~{\sc i} line at $5711.05\,\text{\AA}$, both of which are among the small number of useful GALAH diagnostics for their respective elements. The observations are more than $3\sigma$ deeper than the synthesis in $10\,\mathrm{\%}$ and $13\,\mathrm{\%}$ of the spectra, respectively, while neither line is more than $3\sigma$ shallower in a measurable fraction of the sample. The GALAH DR4 Solar abundances inferred from these lines are correspondingly offset by $+0.171\,\mathrm{dex}$ for Na and $+0.085\,\mathrm{dex}$ for Mg relative to \citet{Grevesse2007}. These results are consistent with underestimated modelled opacity, but do not uniquely establish that the adopted oscillator strengths are responsible. Because the required Solar abundances are themselves offset from the adopted reference scale, the relevant synthetic spectra may also probe a less well covered region of the neural-network training grid used in GALAH DR4 to create them.

The Li~{\sc i} feature near $6707.8\,\text{\AA}$ shows the structured mismatch previously associated with limitations of the GALAH DR4 neural-network synthesis \citep[Sec.~8.3.4]{Buder2025b}. By contrast, the O~{\sc i} triplet at $7771$--$7775\,\text{\AA}$ shows recurrent residuals below $5\,\mathrm{\%}$.

The catalogue additionally contains candidate residual associations with Mn~{\sc i} at $4748.15\,\text{\AA}$, Gd~{\sc ii} near $4791.15\,\text{\AA}$, possible Ti~{\sc i}, Cd~{\sc i}, or W~{\sc i} contributors near $4799.85\,\text{\AA}$, Mo~{\sc i} at $4833.896$ and $5858.30\,\text{\AA}$, Pr~{\sc ii} at $4877.87\,\text{\AA}$, and Sm~{\sc i} at $4883.99\,\text{\AA}$. Several of these regions are blended or ambiguous and should therefore be treated as candidate associations of rare elements with potential for further investigation rather than secure new line identifications.

The 13 \textit{Line profile} cases cannot be explained adequately through a scalar change in line strength. They include the Balmer lines and several asymmetric, split, or wing-dominated residuals associated with possible neglected or inaccurate hyperfine or isotopic structure, isotope shifts, component wavelengths, or imperfect broadening.~{\sc i} by \citet{Jofre2017}, involve Cu~{\sc i} 5700.2, V~{\sc i} 5761.4, as well as Sc~{\sc i} 5671.9 and 5686.9. These cases require improved component patterns or line-formation calculations rather than a simple adjustment of $\log gf$.

The Balmer residuals are the clearest large-scale example. Both $\mathrm{H}\upalpha$ and $\mathrm{H}\upbeta$ show broad residual structure whose amplitude and sign vary with stellar parameters, and both appear among the strongest activity-sensitive coefficients. Their behaviour could reflect several combined effects, including broadening, continuum normalisation, atmospheric structure, departures from LTE, and chromospheric filling. The current analysis establishes that their residual profiles vary systematically, but does not isolate the contribution of each physical mechanism.

Residuals are also present in the left wing of the near-infrared Al~{\sc i} doublet at $7835.3$ and $7836.1\,\text{\AA}$. Broad CCD3 structure further complicates the Eu~{\sc ii} region near $6645\,\text{\AA}$, where Co~{\sc i} and Fe~{\sc i} blends may also be imperfectly modelled. Thus, broad instrumental or reduction structure can coexist with, and partly hide, local stellar-line problems.

The 8 known missing and 38 unidentified features provide a second set of priorities. The known missing transitions can be tested directly for inclusion in future line lists. The unidentified regions may contain omitted atomic or molecular transitions, poorly attributed blends, or residual reduction structure. For these lines, we do not identify strong corresponding signals in the data-driven abundance extraction of Sec.~\ref{sec:analysis_cannon_fitted_abundances}. Possible identifications involving rarely measured species remain especially interesting, but require confirmation from independent atlases, laboratory data, or spectra with different wavelength coverage and resolution -- in particular of stars for which the abundances of the potential elements are significantly enhanced and absorption is more likely.

\subsubsection{Non-stellar and reduction-related residuals}

The classification map also shows that not all recurrent residuals belong to the stellar synthesis. The 12 interstellar regions include established diffuse interstellar bands and interstellar K~{\sc i}. Their appearance in the stellar rest frame depends on the relative stellar and interstellar velocities, which can broaden the overall response or produce asymmetric residuals when an interstellar feature overlaps a fitted photospheric line.

The 15 skyline regions demonstrate an analogous effect for emission fixed in the observed frame. Imperfect sky subtraction leaves positive residual flux whose apparent position is shifted when the spectrum is transformed into the stellar rest frame. Their influence on Model~\textcolor{M4}{M4} illustrates how a small number of spectra can generate an apparent label correlation when observing conditions happen to correlate with the sampled activity distribution.

The broad undulating structure in CCD3 provides another example of a residual that cannot be explained through individual stellar lines, but can complicate the interpretation of genuine lines embedded within the pattern. Its origin is more likely connected to reduction and normalisation, or broad instrumental effects than to a sequence of unrelated atomic transitions.

These cases motivate analysing residual components in their natural frames. Stellar-line residuals are most coherent in the stellar rest frame, interstellar features in an ISM rest frame, and sky or telluric features in the observed frame. A single-frame analysis necessarily broadens or blends some of these components and may conceal their true morphology.

\subsection{Residual information is predictive, but not always directly attributable}
\label{sec:discussion_predictive_information}

The data-driven experiments establish that residual spectra contain predictive information, but also show that the physical interpretation of that information depends on the label, sample, and validation method. Tab.~\ref{tab:recovery_performance} summarises the recovery of the common global labels.

\subsubsection{Information about fitted stellar labels}

The recovery of $T_\mathrm{eff}$, $\log g$, and $[\mathrm{Fe/H}]$ demonstrates that the remaining mismatch changes coherently across the label space used in the original fit. Model~\textcolor{M1}{M1*} confirms this within the more restricted \textit{abundance} sample, although its different signal-to-noise ratio, quality selection, parameter coverage, and label distribution prevent attribution of the improved performance to one factor. Balmer and numerous narrow metal-line residuals contribute to the global labels, whereas Mg and Mn are recovered only modestly and Na and Ba scarcely outperform a central prediction. Their non-element-specific coefficients show that Model~\textcolor{M2}{M2} mainly exploits chemical and population correlations rather than isolated missing opacity.

This distinction matters for transferability. A correlation that predicts an abundance within the GALAH training sample may not generalise to a population with different chemical or stellar-parameter correlations. Predictive recovery should therefore not automatically be interpreted as evidence that the residual model has identified missing opacity from the corresponding element.

\subsubsection{Information about unfitted and correlated properties}

The extinction experiment provides the clearest physical validation for an unfitted label. Although the strongly skewed $A_{K_\mathrm{S}}$ distribution limits the interpretation of the global recovery ratio, the more extinguished sightlines are preferentially assigned larger values. More importantly, the extinction coefficient recovers the strongest known diffuse interstellar bands and interstellar K~{\sc i}. Their independently known wavelengths and morphology demonstrate that the model identifies genuine interstellar absorption remaining outside the stellar synthesis.

The incomplete recovery of weaker DIBs is expected because the current analysis uses individual spectra in the stellar rest frame and is dominated by low-extinction sightlines. Relative stellar and interstellar velocities broaden weak features, whereas dedicated DIB analyses \citep[e.g.][]{Vogrincic2023, Saydjari2023} can shift spectra into an interstellar frame and combine many reddened observations.

The activity experiment provides a contrasting caution. $\mathrm{H}\upalpha$ and $\mathrm{H}\upbeta$ show plausible activity-sensitive responses, and several additional coefficient features coincide with lines previously associated with activity. However, the apparent in-sample recovery of $\log R'_\mathrm{HK}$ does not persist under cross-validation. The activity model also responds to residual skylines, for example around $7795\,\text{\AA}$. The current sample therefore contains genuine chromospheric information, but it is too small and confounded for reliable out-of-sample prediction.

Together, the abundance, extinction, and activity experiments illustrate a hierarchy of interpretation. The extinction response is supported by independently identifiable physical features. The abundance responses are stable but partly indirect. The activity responses combine genuine stellar signatures with demonstrably unrelated observational correlations. Stability, prediction, and physical causality are therefore distinct properties that must be tested separately.

\subsection{Towards modelling the full spectrum}
\label{sec:discussion_full_spectrum}

The disproportionate information retained by the excluded $6\,\mathrm{\%}$ does not imply that these pixels should simply be restored to the stellar fit. They were excluded because they combine unreliable stellar lines and profiles with Balmer, interstellar, sky, and reduction structure. Their recovery instead shows that information-rich and safely modelled are not synonymous. Isolated stellar-line mismatches require revised line data or additional blends, ideally through reproducible tools for constructing and updating atomic and molecular line lists, such as \textsc{linemake} \citep{linemake}. Structured profiles require improved hyperfine, isotopic, wavelength, broadening, or line-formation information; interstellar features require a separate velocity component; and sky, telluric, and reduction residuals require observational rather than stellar corrections.

A practical route forward is therefore to retain physical synthesis as the baseline while modelling well-validated residual corrections conditionally on stellar and observational labels. Such corrections should be restricted to wavelength regions where their behaviour is stable across independent stellar populations, fields, and observing conditions. Cross-validation within a single sample is necessary but not sufficient; transfer tests across nights or surveys would be needed to establish that the corrections are not survey-specific correlations.

The Balmer-line residuals provide a natural test case for connecting calculations of different physical fidelity. GALAH already employs extensive grids of 1D non-LTE departure coefficients to make non-LTE synthesis practical for survey-scale analysis \citep{Amarsi2020}. However, detailed calculations show that 3D non-LTE effects can alter both the inner and outer Balmer-line wings and can produce significant differences in inferred effective temperature relative to 1D calculations \citep{Amarsi2018}. Observed $\mathrm{H}\upalpha$ profile morphology also carries information about stellar properties and additional atmospheric or velocity structure \citep{Bergemann2016,Sayeed2024}; recent reviews summarise the broader progress and remaining challenges of 3D non-LTE stellar spectroscopy \citep{Lind2024,Bergemann2026}. The strong and reproducible dependence of the $\mathrm{H}\upalpha$ and $\mathrm{H}\upbeta$ residuals on $T_\mathrm{eff}$ and $\log g$ therefore identifies these lines as immediate benchmarks for testing whether more realistic calculations explain the observed residual morphology.

Because full 3D non-LTE synthesis remains too expensive for routine optimisation and uncertainty exploration across large label spaces, a practical intermediate approach would be to calculate a sparse but representative grid of paired 1D LTE and 3D non-LTE spectra and train an emulator of their wavelength-dependent difference. The computationally inexpensive 1D LTE framework could then be retained for fitting and uncertainty propagation, while the learned correction approximates the behaviour of the more realistic calculations within the covered parameter range. Such a model should be trained and validated on withheld 3D calculations, applied only within the interpolation domain of the training grid, and developed separately for those elements and wavelength regions for which sufficiently reliable 3D calculations are available. The present residual audit does not demonstrate such a mapping, but identifies the lines and regions where it would be most valuable to test.

A complete description of an observed spectrum requires distinct stellar, interstellar, terrestrial, sky, instrumental, and reduction components. Separating these contributions will be especially important when searching for signals, such as those from planets, that are much smaller than the stellar spectrum itself.

\subsection{Limitations and priorities for future work}
\label{sec:discussion_limitations}

The present audit uses a restricted GALAH sample and therefore does not capture the full temporal and instrumental diversity of the survey. Residual structures associated with particular nights, fields, velocities, or reduction conditions may be over- or under-represented. Extending the analysis across independent observing periods would establish which features are stable throughout GALAH and which are specific to individual data sets.

The peak catalogue also depends on the adopted significance threshold, minimum width, and peak separation. These choices define a reproducible set of prominent residual regions, but not a unique decomposition of every mismatch. Broad structures can contain several local causes. The classification map should therefore be interpreted as an overview of the dominant catalogue features rather than a complete partition of the residual spectrum.

The adopted classifications remain plausibility assessments. The residual sign often determines whether more or less modelled opacity is required, but does not uniquely distinguish oscillator strengths, abundances, blends, broadening, or atmospheric structure. The Solar and Arcturus atlases also represent only two regions of stellar-label space. Confirmation of individual line-data corrections will require targeted synthesis tests, additional benchmark stars, and laboratory information where available.
 
The data-driven models are limited by their training samples, label distributions, and quadratic form. Their performance describes interpolation within the sampled label space and should not be extrapolated to sparsely represented stars. Importantly, successful out-of-sample recovery demonstrates that predictive information is present, whereas failed recovery does not demonstrate its absence: information may remain in the residuals but be inaccessible to the adopted quadratic \textsc{The Cannon} model, the available training sample, or the particular set of labels. Label covariance also permits indirect prediction, while small samples can produce strong correlations with observational artefacts, as demonstrated by the activity model.

As an exploratory test of how the adopted data-driven model affects these conclusions, we repeated the analysis using the observed continuum-normalised flux $f_{\lambda\,\mathrm{obs}}$ instead of the residual flux $\Delta f_\lambda$, while retaining the same labels, samples, wavelength coverage, and cross-validation procedure\footnote{The replica of all analyses figures with the observed instead of residual fluxes can be found via their prefix \texttt{actual\_*} in the \href{https://github.com/svenbuder/residual_cannon}{repository}.}. The observed spectra improve the recovery of some labels: the recovery scales ($\sigma(\Delta \ell)$) decrease from $150$ to $68\,\mathrm{K}$ for $T_\mathrm{eff}$, from $0.086$ to $0.054\,\mathrm{dex}$ for $[\mathrm{Fe/H}]$, and from $0.070$ to $0.059\,\mathrm{dex}$ for $[\mathrm{Mg/Fe}]$. For several other labels the performance is similar, with residual- versus observed-flux recovery scales of $0.14$ versus $0.13\,\mathrm{dex}$ for $[\mathrm{Ba/Fe}]$, $0.058$ versus $0.059\,\mathrm{dex}$ for $[\mathrm{Mn/Fe}]$, $0.084$ versus $0.082\,\mathrm{dex}$ for $[\mathrm{Na/Fe}]$, and $0.11$ versus $0.10$ for $\log R'_\mathrm{HK}$. Unexpectedly, the observed spectra perform worse for $\log g$ ($0.24$ versus $0.19\,\mathrm{dex}$) and $A_{K_\mathrm{S}}$ ($0.059$ versus $0.039\,\mathrm{mag}$).

These mixed results emphasise that the recovery precision of a particular \textsc{The Cannon} model is not a direct measure of the total information contained in its input spectra. The physical synthesis may remove dominant and highly non-linear spectral variation, leaving residual correlations that are easier for the adopted quadratic model to represent, while limited coverage of label space can further restrict what either model can learn, particularly for sparsely sampled abundance extremes. Conversely, poor recovery from the residuals does not demonstrate that information about a label is absent. A full comparison of otherwise identical models trained on the observed, synthetic, and residual spectra, and of more flexible data-driven architectures, would be required to separate information content from model accessibility; such an analysis is beyond the scope of the present residual-focused study.

Immediate priorities include validating the highest-significance line-strength and line-profile candidates, correcting important blends such as the Si~{\sc i}--Rb~{\sc i} region near $7800\,\text{\AA}$, separating stellar, interstellar, and observed-frame residual components, and constructing larger and more representative activity samples. Applying the same framework to independent surveys would further distinguish limitations common to stellar synthesis from those specific to the GALAH line list, instrument, or reduction.

\section{Conclusions}
\label{sec:conclusions}

We introduced a stellar-rest-frame audit of the residuals between 7087 observed GALAH DR4 spectra and their best-fitting 1D LTE and non-LTE syntheses. Rather than behaving as random noise, the residuals contain coherent, recurrent, and signed structure: $4.6$ million of approximately 99 million pixels exceed $3\sigma$, compared with $0.3$ million expected for independent Gaussian noise.

A reproducible peak search identifies 380 residual regions (Fig.~\ref{fig:residual_statistics}). Comparison with stellar atlases, the GALAH line list, interstellar catalogues, and sky measurements associates most with candidate line-strength or profile mismatches, while others trace missing or unidentified lines, interstellar absorption, sky emission, and broad reduction structure (Fig.~\ref{fig:residual_classification}). Their signs indicate whether the total modelled absorption should increase or decrease, although blends, abundance offsets, hyperfine or isotopic structure, broadening, and line formation remain degenerate. The Si~{\sc i}--Rb~{\sc i} blend near $7800\,\text{\AA}$ illustrates how such residuals can directly limit an astrophysically important abundance measurement.

Using \textsc{The Cannon}, we trained data-driven models on the residual flux rather than the observed flux and recovered $T_\mathrm{eff}$ to $80$--$150\,\mathrm{K}$, $\log g$ to $0.14$--$0.19\,\mathrm{dex}$, and $[\mathrm{Fe/H}]$ to $0.059$--$0.086\,\mathrm{dex}$ (Tab.~\ref{tab:recovery_performance}), demonstrating that label-dependent structure remains after the original fit. Mg and Mn retain modest but partly indirect predictive information, whereas Na and Ba scarcely outperform a central prediction and their coefficients do not trace clean element-specific residuals (Fig.~\ref{fig:cannon_linear_coefficients_joint}). Among unfitted properties, the extinction coefficient recovers established diffuse interstellar bands and interstellar K~{\sc i}. $\mathrm{H}\upalpha$ and $\mathrm{H}\upbeta$ retain plausible activity signatures, but $\log R'_\mathrm{HK}$ is not recovered out-of-sample and skyline correlations expose the risk of confusing prediction with causality.

A particularly striking result is that the 793 pixels excluded from the GALAH DR4 optimisation, corresponding to only approximately $6\,\mathrm{\%}$ of the available wavelength range, recover the global stellar parameters almost as well as the complete residual spectrum and better than the remaining 13134 optimisation pixels alone. These regions were excluded because the synthesis or reduction was considered less reliable there; their strong predictive performance shows that they contain a disproportionate fraction of the remaining label-dependent information. This does not imply that they should simply be returned to the original fit. Rather, their information can only be used reliably once their different stellar, interstellar, sky, and reduction-related contributions are modelled explicitly.

In the current survey era, the critical bottleneck for extracting more information from stellar spectra is shifting from data volume and quality towards model fidelity \citep{Xiang2019,Saydjari2023,Buder2025b,Storm2026,Casey2026}. Coherent residuals in features such as the Balmer-line profiles demonstrate the need for more realistic stellar-atmosphere and line-formation calculations, including 3D non-LTE modelling. The prevalence of line-strength mismatches, line-profile discrepancies, and missing or unidentified absorption features shows that complete, experimentally anchored atomic and molecular data are equally essential for realising the precision of large spectroscopic surveys.

Extending our audit across fields, epochs, and observing conditions would separate survey-wide modelling deficiencies from local reduction effects. Future models should treat stellar, interstellar, and observed-frame components separately and may combine targeted line-list revisions with validated data-driven corrections or sparse mappings from 1D to 3D non-LTE calculations.

Residual spectra are therefore not merely quality-control products: they identify where the physical and observational description remains incomplete, where information has not yet been extracted, and where predictive correlations require caution. Residual modelling does not replace physical synthesis, but helps diagnose, validate, and prioritise its improvement. In this sense, \textit{one spectroscopist's trash is another's treasure}: reading between the lines maps the information that tomorrow's models should recover.

\section*{Acknowledgments}

We thank David W. Hogg for an inspiring conference summary talk exactly a decade ago that postulated we need to better model the wealth of existing data, and for a lengthy discussion over coffee that reinforced our motivation to investigate how much information remains in spectral residuals.

We acknowledge the traditional owners of the land on which the ANU stands, the Ngunnawal and Ngambri people. We pay our respects to elders past, and present and are proud to continue their tradition of surveying the night sky and its mysteries to better understand our Universe.

SB acknowledges support from the Australian Research Council under grant number DE240100150. SB acknowledges the use of GitHub Copilot within VSCode to assist with code completion and OpenAI's ChatGPT (GPT-5.5 Thinking) for brainstorming, critical feedback on analysis choices and interpretations, manuscript organisation, and language refinement. All AI-generated suggestions, including code, were critically reviewed, revised where necessary, and validated by the authors before use. All scientific analyses, interpretations, and conclusions presented in this work are those of the authors.

\section*{Software}

\textsc{python} (version 3.12.11) and included its packages
\textsc{astropy} \citep[v. 7.2.0;]{Robitaille2013,PriceWhelan2018},
\textsc{IPython} \citep[v. 8.37.0;]{ipython},
\textsc{matplotlib} \citep[v. 3.10.9;]{matplotlib},
\textsc{NumPy} \citep[v. 2.2.6;]{numpy},
\textsc{scipy} \citep[v. 1.17.1;]{Scipy};
\textsc{sklearn} \citep[v. 1.9.0;]{scikit-learn};
\textsc{topcat} \citep[v. 4.7;]{Taylor2005};
\textsc{thecannon} \citep[v. 0.9.00 from \url{https://github.com/ANU-RSAA/AnniesLasso} maintained by Marc White from ANU RSAA/AITC's Software Group]{Ness2015, Casey2016}.

\section*{Data Availability}

All code to reproduce the analysis and figures can be publicly accessed via \url{https://github.com/svenbuder/residual_cannon}. All data products of GALAH DR4 are publicly available at \url{https://cloud.datacentral.org.au/teamdata/GALAH/public/GALAH_DR4/} as described in \citet{Buder2025b}.

\bibliographystyle{mnras}
\bibliography{bib}

\begin{appendix}

\section{Supplementary residual diagnostics}

\begin{figure*}
    \centering
    \includegraphics[width=\linewidth]{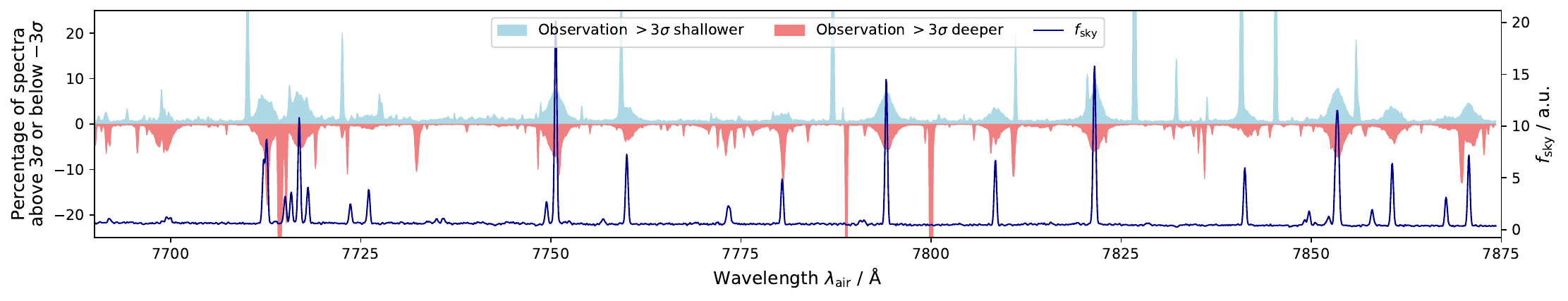}
    \caption{Comparison of the broad residual structure in the fourth GALAH wavelength band with observed sky-emission features. The agreement in wavelength supports the identification of these features as residuals associated with imperfect sky subtraction.}
    \label{fig:skyline_confirmation}
\end{figure*}

\begin{figure*}
    \centering
    \includegraphics[width=\linewidth]{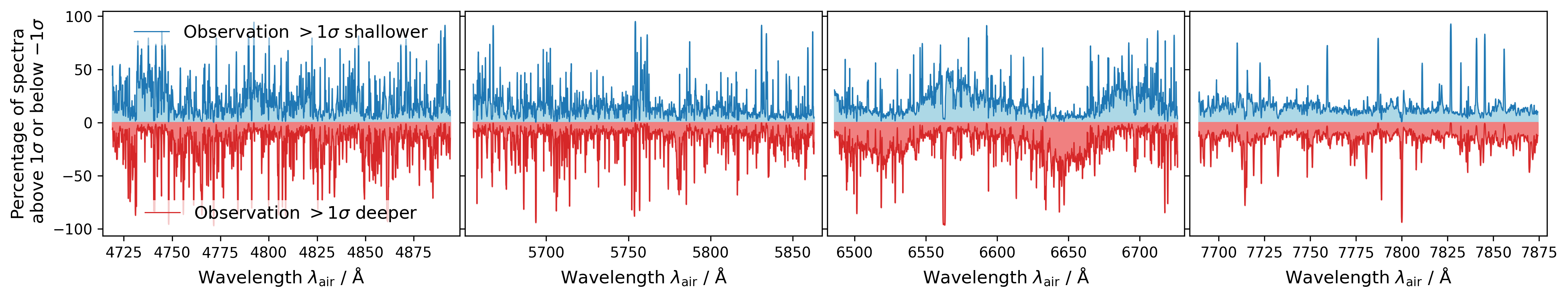}
    \includegraphics[width=\linewidth]{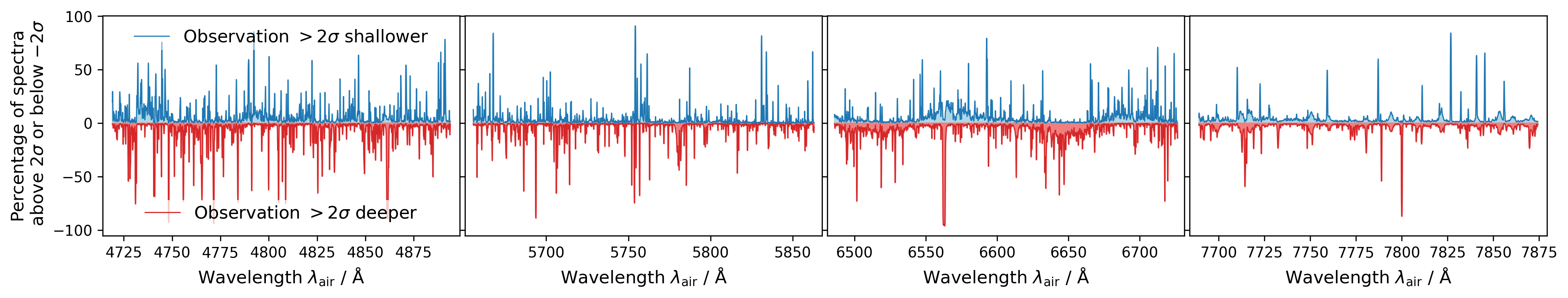}
    \caption{Fraction of spectra with observed flux significantly shallower or deeper than the synthesis at more than $1\sigma$ (top) and $2\sigma$ (bottom), shown across the four GALAH wavelength bands. The lower panels reproduce the significance threshold used in the bottom row of Fig.~\ref{fig:residual_statistics}.}
    \label{fig:percentage_residuals_1_and_2_sigma}
\end{figure*}

\begin{figure*}
    \centering
    \includegraphics[width=\linewidth]{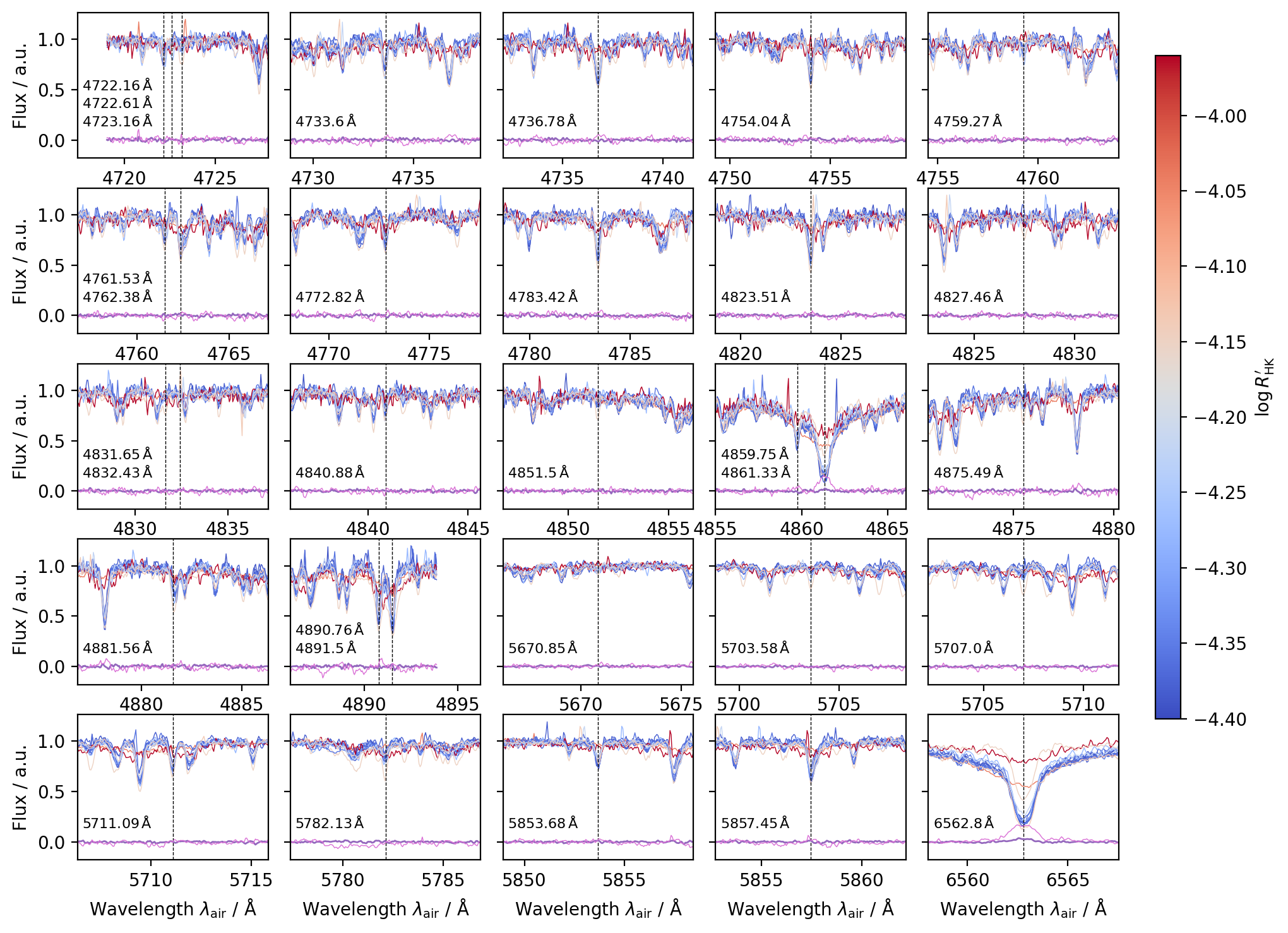}
    \caption{Observed flux for 32 lines listed as activity-sensitive by \citet{Lisogorskyi2019}. Flux is coloured by $\log R'_\mathrm{HK}$ of the \textit{activity} and linear and quadratic coefficients of $\log R'_\mathrm{HK}$ are shown as thick and thin purple lines.}
    \label{fig:logRHK_lines_Lisogorskyi2019}
\end{figure*}

This appendix presents three supporting diagnostics referenced in the main text. Fig.~\ref{fig:skyline_confirmation} compares the broad fourth-arm residual structure with observed sky emission, Fig.~\ref{fig:percentage_residuals_1_and_2_sigma} shows the signed recurrence of residuals at the lower $1\sigma$ and $2\sigma$ thresholds (compared to $3\sigma$ from Fig.~\ref{fig:residual_statistics} bottom), and Fig.~\ref{fig:logRHK_lines_Lisogorskyi2019} shows the residual activity response for lines from \citet{Lisogorskyi2019}.

\end{appendix}

\end{document}